# Pulsar timing array analysis in a Legendre polynomial basis

Bruce Allen [*] and Arian L. von Blanckenburg [†]
*Max Planck Institute for Gravitational Physics (Albert Einstein Institute), Leibniz Universität Hannover, Callinstrasse 38, D-30167, Hannover, Germany*

Ken D. Olum [‡]
*Institute of Cosmology, Department of Physics and Astronomy, Tufts University, Medford, Massachusetts 02155, USA*

(Received 27 October 2025; accepted 5 February 2026; published 4 May 2026)

We use Legendre polynomials, previously employed in this context by Lee *et al.*, van Haasteren and Levin, and Pitrou and Cusin, to model signals in pulsar timing arrays. These replace the (Fourier mode) basis of trigonometric functions normally used for data analysis. The Legendre basis makes it simpler to incorporate pulsar modeling effects, which remove constant-, linear-, and quadratic-in-time terms from pulsar timing residuals. In the Legendre basis, this zeros the amplitudes of the first three Legendre polynomials. We use this basis to construct an optimal quadratic cross-correlation estimator $\hat{\mu}$ of the Hellings and Downs correlation and compute its variance $\sigma^2_{\hat{\mu}}$ in the way described by Allen and Romano. Remarkably, if the gravitational-wave background and pulsar noise power spectra are (sums of) power laws in frequency, then in this basis one obtains analytic closed forms for many quantities of interest.

## I. INTRODUCTION

Pulsar timing array (PTA) collaborations have observed pulse times of arrival (TOAs) for many years. Recently, they reported various levels of evidence for a stochastic gravitational-wave background (GWB) [1–5].

To carry out this inference, PTAs compare the observed TOAs to models which incorporate the effects of GWBs and intrinsic pulsar noise (including both variations in the pulsar spin and in interstellar propagation delays). These are typically modeled as independent stationary and Gaussian stochastic random processes, with power-law spectra. The GWB can be distinguished from intrinsic pulsar noise because the GWB induces a characteristic pattern of interpulsar correlations, first predicted by Hellings and Downs (HD) [6].

The analysis of the TOAs is done in terms of the residuals that remain after subtracting a deterministic model of the pulse arrival times in the absence of a GWB. This so-called "timing model" always includes a constant term which depends upon the pulsar's rotational phase and distance to the Solar System Barycenter (SSB), a linear term determined by the pulsar spin frequency and velocity along the line of sight, and a quadratic term determined by the pulsar's intrinsic spin-down rate and its acceleration along the line of sight. We call these three terms the "universal terms," because there may also be other deterministic terms. For example, if the pulsar is in a binary, the orbital motion of the pulsar around its companion results in a periodically changing propagation path length.

The timing model parameters that characterize these terms are not known independently. Rather, they are determined by fitting to the observed TOAs. Thus, the analysis of the residuals must take into account that this fitting process subtracts away constant, linear, and quadratic effects that arise because of the GWB. (Fitting the timing model also removes the other terms, but here we consider only these three universal terms.)

A zero-mean multivariate Gaussian probability distribution is completely determined by its covariance. Thus, a PTA GWB search needs the covariance matrices (and their inverses) for the timing residuals arising from pulsar intrinsic noise contributions and from the GWB [7]. Both can be calculated from the data models, and in principle this can be done in the same (time) domain as the PTA data are collected. The standard approach is described in [8], but note that it incorrectly assumes that the covariance matrices of stationary processes (for example, $H_{jk}$, defined by Eq. (16) in [9]) are diagonal in the discrete Fourier basis.

[*]Contact author: bruce.allen@aei.mpg.de
[†]Contact author: arian.von.blanckenburg@aei.mpg.de
[‡]Contact author: kdo@cosmos.phy.tufts.edu

For the multivariate Gaussian model, the log of the likelihood is a quadratic form in the data; the inverse of the covariance matrix is the "metric" or "kernel" of this quadratic form. Up to an overall normalization, this likelihood is the probability that any specific set of residuals would occur. However, generating posterior samples with this likelihood is computationally intractable: the large number (hundreds of thousands) of TOAs leads to enormous covariance matrices, and computing their inverses is computationally very expensive. Instead, the analysis uses a small basis (typically $\lesssim 100$ elements) to write the covariance matrix in a low-rank form. Then its inverse and the log likelihood can be computed quickly.

To obtain this low-rank form, most PTA analyses use a Fourier basis whose components are sinusoidal functions of time with oscillation frequencies $j/T$, where $T$ is the total time of observation, and $j$ is a positive integer. In this paper, we explore an alternative basis of Legendre polynomials. The advantage of this basis is that the constant, linear, and quadratic terms in the timing model span the same space as the first three Legendre polynomials. Therefore to remove these timing model terms we simply omit these three polynomials from our basis.

The perturbations in pulse production, propagation, and observation are described in terms of timing residuals. These are denoted by $\mathcal{T}_a(t)$, where $a$ specifies the pulsar and $t$ is the time of observation. In the main body of this paper, the analysis is done in terms of these timing residuals.

A brief outline of the paper follows. In Sec. II, we describe the Fourier basis, and in Sec. III we introduce the Legendre basis. In Sec. IV, we show how we can use it to remove the universal terms in the timing model. In Sec. V, we compute the covariance matrix in the Legendre basis, and in Sec. VI we evaluate this matrix for power-law spectra. Remarkably, the matrix elements have a simple analytic closed form in terms of gamma functions. In Sec. VII, we work out the relationship and connection between the Legendre and Fourier bases. In Sec. VIII, we give the formula for the optimal HD estimator of [9] and its variance, in terms of timing-residual quantities, in the Legendre basis. These are analogous to the corresponding Fourier basis expressions given in [9]. In Sec. IX, we extend our analysis to pulsars that are observed over different time intervals, and in Sec. X, we use our techniques to compute the covariance between timing residuals at times $t$ and $t'$. This is no longer stationary (a function only of $t - t'$) because removing the universal terms in the timing model breaks time stationarity. We also obtain the transmission function previously found numerically in [10] and analytically in [11]. This is followed by a short conclusion and discussion in Sec. XI.

The appendixes contain some technical details and calculations. In Appendix A we explicitly compute the projection which removes the universal timing terms in the Fourier basis. In Appendix B we calculate an integral that we need in Sec. X. Equivalent results are given in [12] (particularly Appendix A). In Appendix C we derive the relationship between an analysis carried out on timing residuals (as in the main body of the paper) and an analysis carried out on redshifts, as used in some of the other literature. Finally, in Appendix D, we prove that the same values are obtained for two estimators of interest, starting from either timing residuals or from redshifts in the Legendre basis. This generalizes a similar result obtained in [9].

## II. FOURIER BASIS

To begin, suppose that the pulsar timing residuals are continuously observed on an interval $t \in [-T/2, T/2]$ for some fixed $T > 0$. In the usual Fourier basis, we represent the timing residuals on this interval by the Fourier series

$$\mathcal{T}_a(t) = \sum_j \mathcal{T}_a^j e^{2\pi i f_j t} \quad \text{for } t \in [-T/2, T/2], \tag{2.1}$$

where $f_j = j/T$ are discrete Fourier frequencies, $T$ is the observation time, $\mathcal{T}_a^j$ are complex expansion coefficients, and $j \in \mathbb{Z}$. Indices $a, b, c, d$ will denote pulsars, and indices $j, k, \ell, m$ will label modes in the Fourier basis.

The Fourier basis would be the preferred choice if we were analyzing periodic signals with period $T$. In that case, the covariance matrix of the Fourier modes, $\langle \mathcal{T}_a^j \mathcal{T}_b^{k*} \rangle$, would be diagonal in $j$ and $k$. That is not the case here, because the GWB and pulsars are observed for a finite time and have components whose frequencies do not correspond to an integer number of cycles within the observation time. See the conclusion of [9] for further discussion of this point.

## III. LEGENDRE BASIS

Since the Fourier basis is not preferred, it makes sense to consider other bases. As shown in [11–13], a basis of Legendre polynomials will simplify the fitting of the timing model.[1] With this as motivation, we expand the timing residuals of pulsar $a$ as

$$\mathcal{T}_a(t) = \sum_{\mu=0}^{\infty} \mathcal{T}_a^\mu P_\mu(2t/T) \quad \text{for } t \in [-T/2, T/2], \tag{3.1}$$

where $P_\mu(z)$ is the Legendre polynomial of order $\mu$, illustrated in Fig. 1. The reader should keep in mind that the coefficients $\mathcal{T}_a^\mu$ have a different meaning and numerical values than in (2.1): they are the components in a Legendre basis rather than in a Fourier basis. To emphasize this, we use indices $\mu$, $\nu$, $\lambda$, $\kappa$ to label components in the Legendre basis. We will see that fitting a pulsar timing model removes the $\mu = 0$, 1, 2 terms from the sum in (3.1) [11–13].

[1] While [13] does not contain the word "Legendre," the first three Legendre polynomials appear in Eq. (A1) in [13] as $\hat{f}_1(t)$, $\hat{f}_2(t)$ and (typo corrected) $\hat{f}_3(t)$. See footnote 2 in Sec. III.

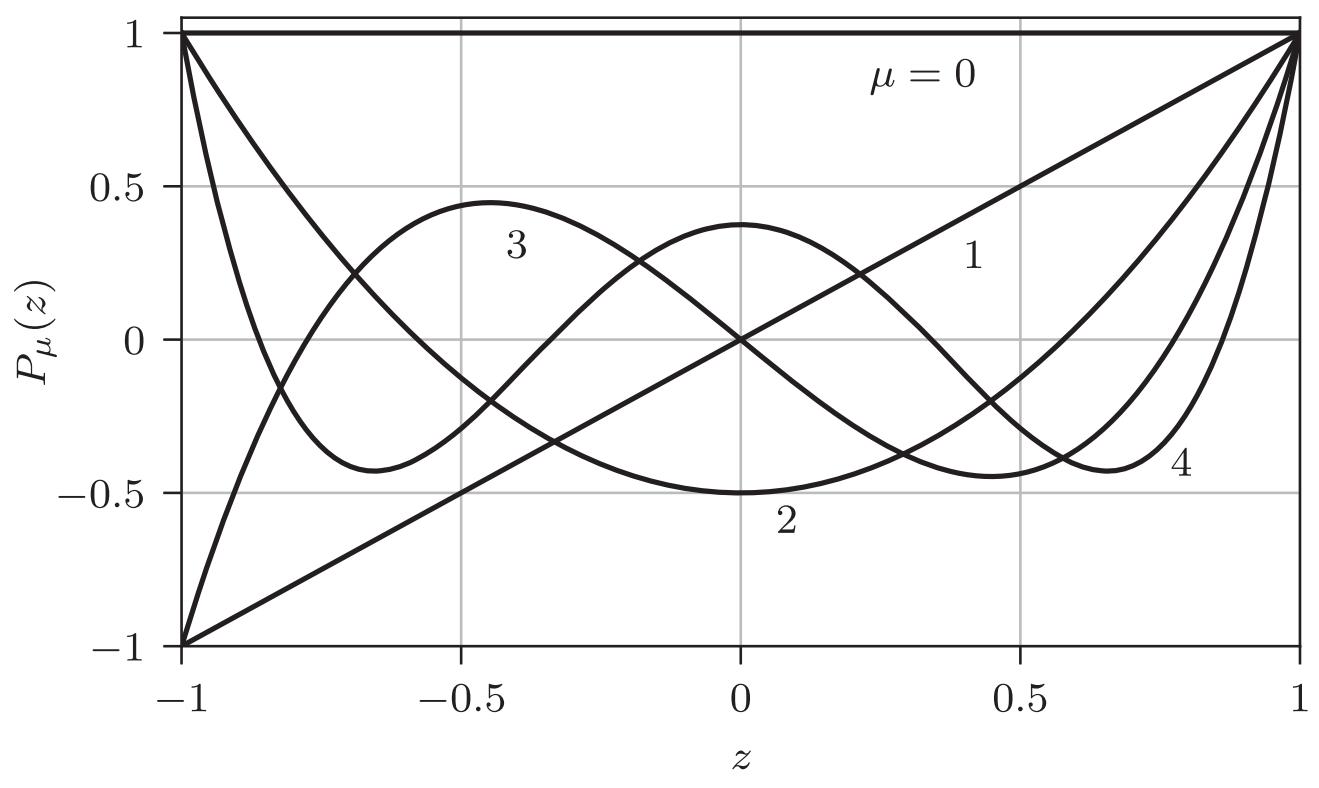

FIG. 1. The Legendre polynomials $P_\mu(z)$ for $\mu = 0, 1, \ldots, 4$.

We use conventional Legendre polynomials, which have normalization

$$\int_{-1}^{1} \mathrm{d}z\, P_\mu(z) P_\nu(z) = \frac{2}{2\mu+1}\delta_{\mu\nu}, \tag{3.2}$$

where $\delta_{\mu\nu}$ is the Kronecker delta. The coefficients which appear in (3.1) can be found by multiplying both sides of (3.1) by $P_\nu(2t/T)$ and integrating over $t \in [-T/2, T/2]$. Using (3.2) to evaluate the integral eliminates all but one term from the sum, giving

$$\mathcal{T}_a^\mu = \frac{2\mu+1}{T}\int_{-T/2}^{T/2} \mathrm{d}t\, P_\mu(2t/T)\mathcal{T}_a(t). \tag{3.3}$$

Since the Legendre polynomials and $\mathcal{T}_a(t)$ are real, the expansion coefficients $\mathcal{T}_a^\mu$ are real.

There are many different families of orthogonal polynomials. Here, the Legendre polynomials are preferred because (1) the corresponding measure is uniform in time, and (2) the power term $z^k$ is a linear combination of Legendre polynomials $P_\mu$ of order $\mu \le k$ and consequently is orthogonal to polynomials of order $\mu > k$.[2] The first condition is relevant because (in fitting parameters via pulsar timing) the postfit timing residuals are weighted uniformly in time. The second condition isolates the effects of pulsar timing model fitting to the first three coefficients in the expansion.

The equivalent of Parseval's theorem is obtained by squaring (3.1) and integrating over $t \in [-T/2, T/2]$. Using (3.2) converts the double sum to a single sum, and dividing both sides by $T$ gives

$$\frac{1}{T}\int_{-T/2}^{T/2} \mathrm{d}t \big[\mathcal{T}_a(t)\big]^2 = \sum_\mu \frac{1}{2\mu+1}(\mathcal{T}_a^\mu)^2. \tag{3.4}$$

Here, and going forward, we often do not explicitly indicate the summation range (here, $\mu = 0, 1, \ldots$).

This relation makes it easy to incorporate the effects of removing constant-, linear-, and quadratic-in-time terms from $\mathcal{T}_a(t)$: we will see that it corresponds to setting $\mathcal{T}_a^\mu = 0$ for $\mu = 0$, 1, 2. Thus, in many places in this paper, the summations over the Legendre indices start at $\mu = 3$.

## IV. EFFECTS OF FITTING A PULSAR TIMING MODEL

Pulsar timing residuals are typically obtained using TEMPO [14], TEMPO2 [15], or PINT [16]. Starting from the observed pulse arrival times, these subtract deterministic terms to obtain the "postfit" timing residuals. The parameters of the deterministic models are selected to minimize the sum of the squares of the residuals, effectively minimizing the lhs of (3.4).

Adjusting the pulsar phase, its intrinsic rotation frequency, and its intrinsic spin-down removes terms constant, linear, and quadratic in $t$ from $\mathcal{T}_a(t)$. This is because the coefficients of these terms are selected to minimize the mean-square postfit timing residuals, given by the lhs of (3.4). The postfit timing residuals are then given by

$$\check{\mathcal{T}}_a(t) \equiv (Q\mathcal{T}_a)(t) \tag{4.1}$$

with $Q$ denoting the linear operator which projects out terms constant, linear, and quadratic in $t$. We use the háček accent ˇ to denote quantities after this operator has been applied.

The beauty of the Legendre polynomial decomposition is that the effects of this fitting process are very simple to describe. Because the lhs of (3.4) is the mean-squared timing residual, and the fitting process minimizes this, it follows from the rhs that removal of the constant term from $\mathcal{T}_a(t)$ sets $\mathcal{T}_a^\mu \to 0$ for $\mu = 0$, removal of the constant and linear terms sets $\mathcal{T}_a^\mu \to 0$ for $\mu = 0$, 1 and removal of the constant, linear, and quadratic terms sets $\mathcal{T}_a^\mu \to 0$ for $\mu = 0$, 1, 2. This is because (a) the Legendre polynomial $P_\mu(2t/T)$ is a polynomial of order $\mu$ in $t$ and (b) Parseval's theorem (3.4) shows that the minimization is achieved by driving the corresponding coefficients to zero. Hence the components of the projection are

[2] One can construct the Legendre polynomials via a Gram-Schmidt orthonormalization procedure on the interval $z \in [-1, 1]$ with inner product $(A, B) = \int_{-1}^{1} \mathrm{d}z A(z)B(z)$. The first basis element is the constant function. The second basis element is the linear function $z$, which is already orthogonal to the first basis element. The third basis element is obtained from the quadratic function $z^2$, after subtracting a constant term to make it orthogonal to the first. And so on, with each element being made orthogonal to its predecessors by subtracting appropriate multiples of those with the same parity under reflection $z \to -z$.

$$Q^{\mu}_{\nu} = \begin{cases} \delta^{\mu}_{\nu} & \text{for } 3 \le \mu \text{ and } 3 \le \nu \\ 0 & \text{otherwise} \end{cases} \tag{4.2}$$

in the Legendre basis where $\delta^{\mu}_{\nu}$ is the Kronecker delta.

Now we consider how the timing residuals in the "short" observation time interval $T$ are related to the underlying stochastic processes. Those processes are stationary over cosmological timescales ($10^9$ yr) which are much longer than the observation time $T$ (decades). Over the long time interval the timing residuals (without any subtractions) may be written as a Fourier integral

$$\mathcal{T}_a(t) = \int \mathrm{d}f\, \tilde{\mathcal{T}}_a(f) \mathrm{e}^{2\pi \mathrm{i} f t}, \tag{4.3}$$

where the integral is over $f \in \mathbb{R}$, and a tilde $\tilde{}$ over a time-domain function denotes its frequency-space equivalent.

Since $\mathcal{T}_a(t)$ is real, the resulting Fourier amplitudes satisfy $\tilde{\mathcal{T}}_a(f) = \tilde{\mathcal{T}}_a{}^*(-f)$. If we substitute (4.3) into (3.3), we obtain

$$\mathcal{T}^{\mu}_a = \frac{2\mu+1}{T} \int \mathrm{d}f\, \tilde{\mathcal{T}}_a(f) \int_{-T/2}^{T/2} \mathrm{d}t\, \mathrm{e}^{2\pi \mathrm{i} f t} P_\mu(2t/T). \tag{4.4}$$

The integral may be evaluated using Eq. (7.243.5) in [17], which gives the Bessel functions

$$\int_{-1}^{1} \mathrm{e}^{i\alpha x} P_\mu(x) \mathrm{d}x = \mathrm{i}^\mu \sqrt{\frac{2\pi}{\alpha}} J_{\mu+1/2}(\alpha) \\ = 2\mathrm{i}^\mu j_\mu(\alpha). \tag{4.5}$$

The first equality assumes that $\alpha > 0$. The second equality follows from the definition of the spherical Bessel function $j_\mu(\alpha)$ and holds for all real $\alpha$.

We evaluate (4.4) using (4.5), obtaining

$$\mathcal{T}^{\mu}_a = (2\mu+1)\mathrm{i}^\mu \int \mathrm{d}f\, \tilde{\mathcal{T}}_a(f) j_\mu(\pi f T). \tag{4.6}$$

This is similar to the corresponding Fourier basis equation given in Eq. (13) of [9], with the cardinal sine function (sinc) replaced by a spherical Bessel function $j$ and with timing residuals instead of redshifts.

## V. COVARIANCE BETWEEN HARMONIC COEFFICIENTS IN THE LEGENDRE BASIS

We can now evaluate the covariance between timing residuals for pulsars $a$ and $b$, as given in Allen and Romano (Eq. (15) in [9]). Under the assumption of a stationary Gaussian ensemble describing an isotropic unpolarized GWB and uncorrelated pulsar noise, we have from Eqs. (14) and (18) in [9] for the Fourier amplitudes of redshifts

$$\langle \tilde{Z}_a(f) \tilde{Z}^*_b(f') \rangle = \left[4\pi H(f)\mu_{ab} + N_a(f)\delta_{ab}\right]\delta(f-f'). \tag{5.1}$$

(Note that in [9], the power spectrum of noise from pulsar $a$ was denoted $P_a(f)$, with the corresponding covariance matrix given by $P^{jk}_a$. In this work, to clearly distinguish these quantities from the Legendre polynomials, we instead use $N_a(f)$ and $N^{jk}_a$.) Here, $\mu_{ab}$ is the HD function evaluated for the pulsar pair $ab$ and is given in Eq. (17) of [9], and $H(f)$ and $N_a(f)$ are the two-sided power spectra of the GWB and noise of pulsar $a$ respectively. Note: $\mu_{aa} = 2/3$.

From (C1) and (4.3) it follows that the Fourier amplitudes of timing residuals $\tilde{\mathcal{T}}_a(f)$ are related to the Fourier amplitudes of redshifts $\tilde{Z}_a(f)$ by

$$\tilde{Z}_a(f) = 2\pi \mathrm{i} f \tilde{\mathcal{T}}_a(f). \tag{5.2}$$

Hence from (5.1) we obtain

$$\langle \tilde{\mathcal{T}}_a(f) \tilde{\mathcal{T}}^*_b(f') \rangle = \frac{\langle \tilde{Z}_a(f) \tilde{Z}^*_b(f') \rangle}{4\pi^2 f f'} \\ = \left[\frac{H(f)}{\pi f^2}\mu_{ab} + \frac{N_a(f)}{4\pi^2 f^2}\delta_{ab}\right]\delta(f-f'). \tag{5.3}$$

Thus, from (4.6) it follows that

$$\langle \mathcal{T}^\mu_a \mathcal{T}^\nu_b \rangle = (2\mu+1)(2\nu+1)\mathrm{i}^{\mu-\nu} \\ \times \int \mathrm{d}f \int \mathrm{d}f' \langle \tilde{\mathcal{T}}_a(f) \tilde{\mathcal{T}}^*_b(f') \rangle j_\mu(\pi f T) j_\nu(\pi f' T) \\ = J^{\mu\nu}\mu_{ab} + M^{\mu\nu}_a \delta_{ab}. \tag{5.4}$$

Here, we have used (5.3) and defined the covariance matrices

$$J^{\mu\nu} \equiv (2\mu+1)(2\nu+1)\mathrm{i}^{\mu-\nu} \int \mathrm{d}f \frac{H(f)}{\pi f^2} j_\mu(\pi f T) j_\nu(\pi f T) \\ M^{\mu\nu}_a \equiv (2\mu+1)(2\nu+1)\mathrm{i}^{\mu-\nu} \int \mathrm{d}f \frac{N_a(f)}{4\pi^2 f^2} j_\mu(\pi f T) j_\nu(\pi f T). \tag{5.5}$$

These expressions are very similar to Eq. (16) in Allen and Romano [9], but the sinc (cardinal sine) functions which appear there are replaced here by the (closely related) spherical Bessel functions $j_\mu$. The other difference from [9] is that we use upper indices for the components $J^{\mu\nu}$.

Note that the lhs of Eq. (15) in [9], with redshifts replaced by timing residuals, is the Fourier basis equivalent of (5.4) and contains a complex conjugate $Z^{k*}_b$. Here, there is no complex conjugate, because $\mathcal{T}^\mu_b$ is real.

We can write these expressions as single-sided integrals. Indeed, since $j_\mu(-x) = (-1)^\mu j_\mu(x)$ and the power spectra are even functions $H(f) = H(-f)$ and $N_a(f) = N_a(-f)$, the rhs of (5.5) vanishes if $\mu + \nu$ is odd (which also implies that $\mu - \nu$ is odd). Thus,

$$J^{\mu\nu} = \begin{cases} 0 & \text{for } \mu+\nu \text{ odd} \\ 2(2\mu+1)(2\nu+1)(-1)^{\frac{\mu-\nu}{2}} \int_0^\infty \mathrm{d}f \, \frac{H(f)}{\pi f^2} j_\mu(\pi f T) j_\nu(\pi f T) & \text{for } \mu+\nu \text{ even} \end{cases}$$

$$M_a^{\mu\nu} = \begin{cases} 0 & \text{for } \mu+\nu \text{ odd} \\ 2(2\mu+1)(2\nu+1)(-1)^{\frac{\mu-\nu}{2}} \int_0^\infty \mathrm{d}f \, \frac{N_a(f)}{4\pi^2 f^2} j_\mu(\pi f T) j_\nu(\pi f T) & \text{for } \mu+\nu \text{ even,} \end{cases} \tag{5.6}$$

noting that if $\mu+\nu$ is even then $\mu-\nu$ is also even. In the next section we will evaluate these in closed form for power-law power spectra.

## VI. EVALUATION OF COVARIANCE MATRIX FOR POWER-LAW POWER SPECTRA

The integrals in (5.6) can be computed in closed form for power-law power spectra. For a fixed $\gamma$, take $H(f)$ to be the power law defined by $H(f) \propto |f|^{-\gamma+2}$ (the offset in $\gamma$ is to match standard conventions). For example, the incoherent GWB produced by many independent randomly oriented circular-orbit compact binary systems which are inspiraling due to GW backreaction has a power-law spectrum with $\gamma = 13/3$.

We can see that the integrals in (5.6) converge for a range of $\gamma$ that is large enough to cover the values of interest. The asymptotic behavior of the spherical Bessel functions for small $|z|$ is

$$j_\mu(z) \approx \frac{z^\mu}{(2\mu+1)!!}. \tag{6.1}$$

Since projection (4.2) ensures that $3 \le \mu$ and $3 \le \nu$, the integrals in (5.6) converge as $f \to 0$, provided that $\gamma < 7$. For large $|z|$ the spherical Bessel functions fall off as

$$j_\mu(z) \approx \begin{cases} (-1)^{\frac{\mu}{2}} \left(\frac{\sin z}{z}\right) & \text{for } \mu \text{ even,} \\ (-1)^{\frac{\mu+1}{2}} \left(\frac{\cos z}{z}\right) & \text{for } \mu \text{ odd.} \end{cases} \tag{6.2}$$

Hence, provided that $-1 < \gamma$, the integrals converge as $f \to \infty$. Taking into account the $f \to 0, \infty$ behaviors, the integrals in (5.6) converge, provided that $-1 < \gamma < 7$. This is broad enough to include the entire range of standard GWB and pulsar noise models, including white pulsar noise.

Remarkably, the integrals in (5.6) can be evaluated in closed form for (two-sided) power-law spectra

$$H(f) = \frac{3\pi}{4} \frac{A_{\rm gw}^2}{12\pi^2} \left|\frac{f}{f_{\rm r}}\right|^{-\gamma_{\rm gw}+2} f_{\rm r}^{-1} \tag{6.3}$$

and

$$N_a(f) = 2\pi^2 \frac{A_{\rm rn}^2}{12\pi^2} \left|\frac{f}{f_{\rm r}}\right|^{-\gamma_{\rm rn}+2} f_{\rm r}^{-1}. \tag{6.4}$$

Here $f_{\rm r} > 0$ is a reference frequency, often taken to be 1/yr, and the $A^2$ are dimensionless. The prefactors $3\pi/4$ and $2\pi^2$ in (6.3) and (6.4) are chosen to match standard conventions for the GWB (subscript "gw") and for pulsar red noise (subscript "rn"). To see this, consider the case $a = b$. From Eqs. (2) and (17) in [9] we have $\mu_{aa} = 2/3$. Then, (5.3) reads

$$\begin{aligned} \langle \tilde{\mathcal{T}}_a(f) \tilde{\mathcal{T}}_a^*(f') \rangle &= \left[\frac{2}{3\pi} \frac{H(f)}{f^2} + \frac{N_a(f)}{4\pi^2 f^2}\right] \delta(f-f') \\ &= \frac{1}{2} \frac{A_{\rm gw}^2}{12\pi^2} \left|\frac{f}{f_{\rm r}}\right|^{-\gamma_{\rm gw}} f_{\rm r}^{-3} \delta(f-f') \\ &\quad + \frac{1}{2} \frac{A_{\rm rn}^2}{12\pi^2} \left|\frac{f}{f_{\rm r}}\right|^{-\gamma_{\rm rn}} f_{\rm r}^{-3} \delta(f-f'), \end{aligned} \tag{6.5}$$

where to obtain the second equality we used (6.3) and (6.4). The final equality in (6.5) displays the PTA community's standard amplitude and slope conventions for two-sided power spectral densities of timing residuals induced by GWB and pulsar noise.

We can evaluate the integral for $J^{\mu\nu}$ in (5.6) if $-1 < \gamma_{\rm gw} < 7$, which covers all required slopes. $J^{\mu\nu}$ vanishes if $\mu+\nu$ is odd. For $\mu+\nu$ even, one has

$$\begin{aligned} J^{\mu\nu} &= 2(2\mu+1)(2\nu+1)(-1)^{\frac{\mu-\nu}{2}} \int_0^\infty \mathrm{d}f \, \frac{H(f)}{\pi f^2} j_\mu(\pi f T) j_\nu(\pi f T) \\ &= (2\mu+1)(2\nu+1)(-1)^{\frac{\mu-\nu}{2}} \frac{A_{\rm gw}^2}{16\pi^2 f_{\rm r}^3 T} (\pi f_{\rm r} T)^{\gamma_{\rm gw}} \int_0^\infty \mathrm{d}z \, z^{-\gamma_{\rm gw}-1} J_{\mu+1/2}(z) J_{\nu+1/2}(z) \\ &= (2\mu+1)(2\nu+1)(-1)^{\frac{\mu-\nu}{2}} \frac{A_{\rm gw}^2}{32\pi^2 f_{\rm r}^3 T} \left(\frac{\pi f_{\rm r} T}{2}\right)^{\gamma_{\rm gw}} \frac{\Gamma(\gamma_{\rm gw}+1)\Gamma(\frac{\mu+\nu-\gamma_{\rm gw}+1}{2})}{\Gamma(\frac{\mu-\nu+\gamma_{\rm gw}+2}{2})\Gamma(\frac{\mu+\nu+\gamma_{\rm gw}+3}{2})\Gamma(\frac{\nu-\mu+\gamma_{\rm gw}+2}{2})}. \end{aligned} \tag{6.6}$$

The first equality follows from (5.6), the second from substituting (6.3), changing variables to $z = \pi f T$, and using $j_\mu(z) = \sqrt{\pi/2z} J_{\mu+1/2}(z)$, and the third from Eq. (6.574.2) in [17]. We obtain the integral for $M_a^{\mu\nu}$ in (5.6) by multiplying the rhs of (6.6) by 2/3 and replacing $A_{\rm gw}$ by $A_{\rm rn}$ and $\gamma_{\rm gw}$ by $\gamma_{\rm rn}$. The diagonal values $J^{\mu\mu}$ may be found in Eqs. (A10) and (A11) in [12] after correcting the typo via $m \to n$.

Note that if $\gamma_{\rm gw}$ is an even integer, and $|\mu - \nu| \geq \gamma_{\rm gw} + 2$, then $J^{\mu\nu}$ vanishes, because one of the Γ-functions in the denominator of the rhs of (6.6) is singular. In this case, the matrix $J^{\mu\nu}$ is band diagonal.

Equation (6.6) gives an explicit closed-form expression for the covariance matrix elements for a power-law spectrum in the Legendre basis. Thus, since $J^{\mu\nu}$ and $M_a^{\mu\nu}$ are linear in the frequency domain power spectra, (6.6) provides an explicit covariance matrix for any noise model which is a sum of power laws with slopes $\gamma$ satisfying $-1 < \gamma < 7$. Note that for $\mu,\ \nu \geq 3$, (6.6) is finite for $-1 < \gamma_{\rm gw} < 7$, with singularities at $\gamma_{\rm gw} = -1, 7$, consistent with our convergence discussions at the start of this section.

## VII. RELATION BETWEEN FOURIER BASIS AND LEGENDRE BASIS

When we express a function with respect to two different bases, for example, timing residuals in (2.1) and (3.1), there exists a linear transformation between the components in the different bases. Suppose that we have the values of the Fourier basis components $\mathcal{T}_a^k$ and that we want to transform them into the Legendre basis. The transformation is defined by quantities $\mathbf{\Lambda} \equiv \Lambda^\mu{}_k$ with components

$$\Lambda^\mu{}_k \equiv (2\mu + 1) \mathrm{i}^\mu j_\mu(\pi k). \tag{7.1}$$

This transformation is obtained by setting (2.1) equal to (3.1), multiplying both sides by $P_\nu(2t/T)$ and integrating over $t \in [-T/2, T/2]$, using (3.2) and (4.5).

In what follows, we use the Einstein summation convention: if the same indices appear on different levels in a product, then they are summed over their range. With this convention, the transformation from Fourier coefficients $\mathcal{T}_a^k$ to Legendre coefficients $\mathcal{T}_a^\mu$ is given by

$$\mathcal{T}_a^\mu = \Lambda^\mu{}_k \mathcal{T}_a^k. \tag{7.2}$$

We define the components of the adjoint $\mathbf{\Lambda}^\dagger$ of $\mathbf{\Lambda}$ by

$$(\Lambda^\dagger)_k{}^\mu \equiv (\Lambda^\mu{}_k)^*. \tag{7.3}$$

The inverse transformation $\mathbf{\Lambda}^{-1} \equiv (\Lambda^{-1})^k{}_\mu$ of $\mathbf{\Lambda}$ is defined by

$$\begin{aligned} (\mathbf{\Lambda}^{-1}\mathbf{\Lambda})^j{}_k &= (\Lambda^{-1})^j{}_\mu \Lambda^\mu{}_k = \delta^j_k, \\ (\mathbf{\Lambda}\mathbf{\Lambda}^{-1})^\mu{}_\nu &= \Lambda^\mu{}_k (\Lambda^{-1})^k{}_\nu = \delta^\mu_\nu, \end{aligned} \tag{7.4}$$

and the transformation from Legendre coefficients to Fourier coefficients is given by

$$\mathcal{T}_a^k = (\Lambda^{-1})^k{}_\mu \mathcal{T}_a^\mu. \tag{7.5}$$

Using completeness relations of Fourier and Legendre basis functions, one can show by direct computation that the components of $\mathbf{\Lambda}^{-1}$ are

$$(\Lambda^{-1})^j{}_\mu = \frac{1}{2\mu + 1} \delta^{jk} (\Lambda^\dagger)_k{}^\nu \delta_{\nu\mu}. \tag{7.6}$$

Note that $\mathbf{\Lambda}$ is only invertible if we include an infinite number of components. Thus, the implied summations are over the infinite set of positive integers for the Legendre basis indices $(\mu, \nu, \cdots)$ and over the infinite set of all integers for the Fourier basis indices $(j, k, \cdots)$. In practice (see remarks at the start of Appendix D) the Fourier sum is taken in the order $0; -1, 1; -2, 2; \ldots$ and the Legendre sum in the order $0,1,2,3\cdots$ or $3,4,5,\cdots$.

## VIII. THE OPTIMAL HD ESTIMATOR

We assume that the reader is familiar with [9]. There, starting with pulsar redshifts and working in a Fourier basis, the authors derive an optimal estimator of the HD correlation and compute its variance. As detailed in footnote 19 of [9], the method could also be applied to timing residuals by replacing $H(f) \mapsto H(f)/4\pi^2 f^2$ and $N_a(f) \mapsto N_a(f)/4\pi^2 f^2$. In this section, we carry out that computation in the Legendre basis, starting from timing residuals rather than redshifts.

In this work, the quantities $J^{jk}$ and $M_a^{jk}$ are the timing-residual equivalents of $H_{jk}$ and $N_a^{jk}$ as defined in Eq. (16) of [9]. (Note that $N_a^{jk}$ is denoted by $P_a^{jk}$ in [9].) We define

$$\begin{aligned} J^{jk} &\equiv \int \mathrm{d}f \frac{H(f)}{\pi f^2} \mathrm{sinc}\big(\pi(f - f_j)T\big) \mathrm{sinc}\big(\pi(f - f_k)T\big) \\ M_a^{jk} &\equiv \int \mathrm{d}f \frac{N_a(f)}{4\pi^2 f^2} \mathrm{sinc}\big(\pi(f - f_j)T\big) \mathrm{sinc}\big(\pi(f - f_k)T\big), \end{aligned} \tag{8.1}$$

where $f_j \equiv j/T$ and we have corrected the erroneous factor of $4\pi$ which appears in the second line of Eq. (16) in [9]. The reader may find it helpful to compare (5.4) and (5.5) to Eqs. (15) and (16) of [9].

Here, we denote the Fourier basis components of the timing-residual covariances by $D_{ab,cd}^{jk,\ell m}$; they are defined by Eqs. (19) and (20) in [9] with $H_{jk}$ replaced by $J^{jk}$ and $N_a^{jk}$ replaced by $M_a^{jk}$. These are the analog of the redshift covariance matrix components $C_{ab,cd}^{jk,\ell m}$ defined in [9].

We follow the derivation given in [9], but applied to timing residuals. We compute the variance $\sigma_{\hat{\mu}}^2$ of the optimal estimator of the HD correlation for an angular bin at pulsar sky separation angle $\gamma$. The form of the result is identical to Eqs. (30)–(32) in [9]:

$$\sigma^2_{\hat{\mu}} = \frac{\mu^2_{\rm u}(\gamma)}{(\boldsymbol{\mu}\bar{\boldsymbol{J}})^t \boldsymbol{D}^{-1}(\boldsymbol{\mu}\bar{\boldsymbol{J}})}, \tag{8.2}$$

where the components of $\bar{\boldsymbol{J}}$ in the Fourier basis are $\bar{J}^{jk} \equiv J^{j,-k}$, and the HD vector $\boldsymbol{\mu} = \mu_{ab}$ contains only those pulsar pairs $ab \in \gamma$ in the angular bin at angle $\gamma$.

In [9] the authors also calculate the square of the expected signal-to-noise-ratio (SNR) $\rho$ for that angular bin, in the presence of a GWB. The corresponding timing-residual equivalent of Eq. (38) in [9] is

$$\rho^2 = (\boldsymbol{\mu}\bar{\boldsymbol{J}})^t \boldsymbol{D}_0^{-1}(\boldsymbol{\mu}\bar{\boldsymbol{J}}), \tag{8.3}$$

where $\boldsymbol{D}_0$ is the covariance matrix obtained by setting $\boldsymbol{J} \to 0$ in $\boldsymbol{D}$.

The same expression (8.3) *also* gives the conventional optimal cross-correlation "deflection" statistic (squared SNR) [7]. To obtain that from (8.3), one constructs the vector $\boldsymbol{\mu}$ to contain *all* distinct pulsar pairs, rather than only those in some angular bin.

These calculations and these resulting formulas do not take into account the effects of fitting to intrinsic and extrinsic pulsar parameters, which takes place in normal PTA analysis. As previously discussed, this fitting removes terms from the pulsar timing residuals, including the "universal" terms. In passing, [9] comments that the effects of such fitting and removal can be "gracefully accommodated" by the formalism,[3] but no details are provided. In the remainder of this section, we explicitly show how this fitting and removal affects the calculations and formulas.

### A. Variance $\sigma^2_{\hat{\mu}}$ and SNR $\rho^2$ in the Legendre basis

The variance of the reconstructed HD correlation $\sigma^2_{\hat{\mu}}$ and its associated squared SNR $\rho^2$ are scalars, which are independent of the basis in which they are computed. Equations (8.2) and (8.3) give them in a Fourier basis. Here, we express the same quantities in the Legendre basis and show explicitly that the result is the same in the limit of an infinite number of modes.

By inserting (7.2) into (5.5) and using (7.3), the components $J^{\mu\nu}$ and $M_a^{\mu\nu}$ in the Legendre basis are related to the components $J^{jk}$ and $M_a^{jk}$ in the Fourier basis (8.1) by

$$\begin{aligned} J^{\mu\nu} &= \Lambda^\mu{}_j J^{jk} (\Lambda^\dagger)_k{}^\nu \\ M_a^{\mu\nu} &= \Lambda^\mu{}_j M_a^{jk} (\Lambda^\dagger)_k{}^\nu. \end{aligned} \tag{8.4}$$

From (8.4) it follows that the components $D^{jk,\ell m}_{ab,cd}$ of the timing-residual covariance matrix in the Fourier basis are transformed to the Legendre basis by

$$D^{\mu\nu,\lambda\kappa}_{ab,cd} = \Lambda^\mu{}_j \Lambda^\nu{}_k D^{jk,\ell m}_{ab,cd} (\Lambda^\dagger)_\ell{}^\lambda (\Lambda^\dagger)_m{}^\kappa. \tag{8.5}$$

An analogous formula holds for the components of $\boldsymbol{D}_0$.

The components of the pseudoinverse $\boldsymbol{D}^{-1}$ are transformed to the Legendre basis by

$$(D^{-1})^{ab,cd}_{\mu\nu,\lambda\kappa} = (\Lambda^{-\dagger})_\mu{}^j (\Lambda^{-\dagger})_\nu{}^k (D^{-1})^{ab,cd}_{jk,\ell m} (\Lambda^{-1})^\ell{}_\lambda (\Lambda^{-1})^m{}_\kappa, \tag{8.6}$$

where $\boldsymbol{\Lambda}^{-\dagger} \equiv (\boldsymbol{\Lambda}^{-1})^\dagger$. Unlike [9], for notational consistency we swap upper and lower indices of the components of $\boldsymbol{D}^{-1}$.

The transformation of the Fourier components $\bar{J}^{jk} \equiv J^{j,-k}$ of $\bar{\boldsymbol{J}}$ to the Legendre basis requires some care: "negative indices" have no Legendre-basis equivalent. However, requiring that (8.2) and (8.3) stay invariant under transformations (7.2), (8.4), (8.5), and (8.6) to the Legendre basis, implies that

$$\bar{J}^{\mu\nu} \equiv J^{\mu\nu}. \tag{8.7}$$

This brings us to a simple conclusion. After carrying out the replacements

$$\left\{ \begin{matrix} J^{jk} \\ M_a^{jk} \\ \bar{J}^{jk} \\ (D^{-1})^{ab,cd}_{jk,\ell m} \end{matrix} \right\} \mapsto \left\{ \begin{matrix} J^{\mu\nu} \\ M_a^{\mu\nu} \\ J^{\mu\nu} \\ (D^{-1})^{ab,cd}_{\mu\nu,\lambda\kappa} \end{matrix} \right\}, \tag{8.8}$$

the formulas (8.2) and (8.3) still hold. Indeed, with suitable notational changes, *all* of the equations in [9] still hold in the Legendre basis. The only changes needed are that the "frequency bin" indices $j, k, \ell$, and $m$ are replaced by "Legendre polynomial" indices $\mu$, $\nu$, $\lambda$, and $\kappa$. For the variance and SNR, we have

$$\sigma^2_{\hat{\mu}} = \frac{\mu^2_{\rm u}(\gamma)}{\sum_{ab\in\gamma} \sum_{cd\in\gamma} \sum_{\mu,\nu,\lambda,\kappa} \mu_{ab} J^{\mu\nu} (D^{-1})^{ab,cd}_{\mu\nu,\lambda\kappa} \mu_{cd} J^{\lambda\kappa}}, \tag{8.9}$$

and

$$\rho^2 = \sum_{ab\in\gamma} \sum_{cd\in\gamma} \sum_{\mu,\nu,\lambda,\kappa} \mu_{ab} J^{\mu\nu} (D_0^{-1})^{ab,cd}_{\mu\nu,\lambda\kappa} \mu_{cd} J^{\lambda\kappa}. \tag{8.10}$$

In these formulas, the Legendre component indices $\mu$, $\nu$, $\lambda$, and $\kappa$ are summed over all positive integers.

Writing $\sigma^2_{\hat{\mu}}$ and $\rho^2$ in the Legendre basis now makes it easy to account for the effects of removing the constant, linear, and quadratic terms in the timing residuals. The effect of such removal is simple: In (8.9) and (8.10) we have to replace

$$\begin{aligned} J^{\mu\nu} &\mapsto \check{J}^{\mu\nu} \equiv Q^\mu_\lambda J^{\lambda\kappa} Q^\nu_\kappa, \\ D^{\mu\nu,\lambda\kappa}_{ab,cd} &\mapsto \check{D}^{\mu\nu,\lambda\kappa}_{ab,cd} \equiv Q^\mu_\alpha Q^\nu_\beta D^{\alpha\beta,\gamma\delta}_{ab,cd} Q^\lambda_\gamma Q^\kappa_\delta. \end{aligned} \tag{8.11}$$

[3] See paragraph between Eqs. (12) and (13) in [9].

Note that taking the pseudoinverse and projecting are operations that do not commute if applied to nondiagonal $\boldsymbol{D}$. The correct order here is to first project $\boldsymbol{D}$ to $\check{\boldsymbol{D}}$ and then use the pseudoinverse of $\check{\boldsymbol{D}}$ (or $\check{\boldsymbol{D}}_0$) in (8.9) and (8.10). If $a < b$ and $e < f$, the pseudoinverse $\check{\boldsymbol{D}}^{-1}$ satisfies

$$\sum_{cd\in\gamma}\sum_{\lambda\kappa}(\check{D}^{-1})^{ab,cd}_{\mu\nu,\lambda\kappa}\check{D}^{\lambda\kappa,\alpha\beta}_{cd,ef} = \delta^a_e\delta^b_f Q^\alpha_\mu Q^\beta_\nu. \tag{8.12}$$

In Appendix A we discuss how to account for the timing model subtractions in the Fourier basis.

## IX. DIFFERING PULSAR OBSERVATION TIMES

Up to this point, we have assumed that all pulsars are observed over the same time interval $t\in[-T/2, T/2]$. However, this is not the case for most PTAs, which are composed of pulsars that have different observational time spans. In this section, we modify our calculations to account for this.

Suppose that each pulsar is observed over a different time interval. Let $T_a$ denote the total observing time span for pulsar $a$, and let $\Delta_a$ denote the time at the center of that interval. So the pulsar $a$ observation interval is

$$\Delta_a - \frac{T_a}{2} \le t \le \Delta_a + \frac{T_a}{2}. \tag{9.1}$$

Our previous calculations correspond to the case where $T_a = T$ and $\Delta_a = 0$ for all pulsars.

In the time interval (9.1), each pulsar's timing residuals can be expanded in Legendre polynomials, analogous to (3.1). Thus,

$$\mathcal{T}_a(t) = \sum_\mu \mathcal{T}^\mu_a P_\mu\big(2(t-\Delta_a)/T_a\big) \quad \text{for } t\in\left[\Delta_a - \frac{T_a}{2}, \Delta_a + \frac{T_a}{2}\right]. \tag{9.2}$$

The real coefficients $\mathcal{T}^\mu_a$ are obtained in the same way as for (3.3), yielding

$$\mathcal{T}^\mu_a = \frac{2\mu+1}{T_a}\int_{-T_a/2}^{T_a/2}\mathrm{d}t\, P_\mu(2t/T_a)\mathcal{T}_a(t+\Delta_a). \tag{9.3}$$

Substituting the Fourier integral (4.3) into (9.3) and carrying out the integral with (4.5) gives the Legendre basis coefficients in term of the Fourier amplitudes, as

$$\mathcal{T}^\mu_a = (2\mu+1)\mathrm{i}^\mu\int \mathrm{d}f\, \tilde{\mathcal{T}}_a(f) j_\mu(\pi f T_a)\mathrm{e}^{2\pi\mathrm{i}f\Delta_a}. \tag{9.4}$$

This correctly reduces to (4.6) in the limit $\Delta_a \to 0$ and $T_a \to T$.

The covariance of these Legendre coefficients is still given by (5.4), $\langle \mathcal{T}^\mu_a\mathcal{T}^\nu_b\rangle = J^{\mu\nu}\mu_{ab} + M^{\mu\nu}_a\delta_{ab}$, but the covariance matrices are now

$$J^{\mu\nu} = (2\mu+1)(2\nu+1)\mathrm{i}^{\mu-\nu}\int \mathrm{d}f\frac{H(f)}{\pi f^2}j_\mu(\pi f T_a)j_\nu(\pi f T_b)\mathrm{e}^{2\pi\mathrm{i}f(\Delta_a-\Delta_b)}$$
$$M^{\mu\nu}_a = (2\mu+1)(2\nu+1)\mathrm{i}^{\mu-\nu}\int \mathrm{d}f\frac{N_a(f)}{4\pi^2 f^2}j_\mu(\pi f T_a)j_\nu(\pi f T_b)\mathrm{e}^{2\pi\mathrm{i}f(\Delta_a-\Delta_b)}. \tag{9.5}$$

Note that these matrices are real, but are no longer symmetric in $\mu$ and $\nu$.

To write these as single-sided integrals, first consider $J^{\mu\nu}$. For $\mu+\nu$ even (implying $\mu-\nu$ even) the quantity $H(f)j_\mu(\pi f T_a)j_\nu(\pi f T_b)/\pi f^2$ is an even function of $f$, so only the real part of the complex phase will contribute. If $\mu+\nu$ is odd (implying $\mu-\nu$ odd) then only the imaginary part contributes. So, we obtain

$$J^{\mu\nu} = \begin{cases} 2(2\mu+1)(2\nu+1)(-1)^{\frac{\mu-\nu+1}{2}}\displaystyle\int_0^\infty \mathrm{d}f\frac{H(f)}{\pi f^2}j_\mu(\pi f T_a)j_\nu(\pi f T_b)\sin\big(2\pi f(\Delta_a-\Delta_b)\big) & \text{for } \mu+\nu \text{ odd} \\ 2(2\mu+1)(2\nu+1)(-1)^{\frac{\mu-\nu}{2}}\displaystyle\int_0^\infty \mathrm{d}f\frac{H(f)}{\pi f^2}j_\mu(\pi f T_a)j_\nu(\pi f T_b)\cos\big(2\pi f(\Delta_a-\Delta_b)\big) & \text{for } \mu+\nu \text{ even.} \end{cases} \tag{9.6}$$

The corresponding equations for $M^{\mu\nu}_a$ are obtained by dividing by $4\pi$ and replacing $H(f)$ with $N_a(f)$.

We now evaluate these integrals for the power-law form of $H(f)$ given in (6.3). In the case where $|\Delta_a - \Delta_b|$ is small compared to $T_a$ and $T_b$, we can use Eq. (6.574.1) from [17] to express these in terms of the Gauss hypergeometric function ${}_2F_1(a,b;d;z)$. In the case where $|\Delta_a - \Delta_b|$ is not small, we can make use of Eq. (6.578.1) in [17]

$$\int_0^\infty x^{\rho-1}J_\lambda(ax)J_\mu(bx)J_\nu(cx)\mathrm{d}x = \frac{2^{\rho-1}a^\lambda b^\mu c^{-\lambda-\mu-\rho}\Gamma(\frac{\lambda+\mu+\nu+\rho}{2})}{\Gamma(\lambda+1)\Gamma(\mu+1)\Gamma(1-\frac{\lambda+\mu-\nu+\rho}{2})}F_4\left(\frac{\lambda+\mu-\nu+\rho}{2},\frac{\lambda+\mu+\nu+\rho}{2};\lambda+1,\mu+1;\frac{a^2}{c^2},\frac{b^2}{c^2}\right)$$
$$\text{assuming } \Re(\lambda+\mu+\nu+\rho)>0,\quad \Re(\rho)<\frac{5}{2},\quad a>0,\quad b>0,\quad c>0,\ \text{and}\ c>a+b, \tag{9.7}$$

where $F_4$ denotes the Appell hypergeometric function. A notable feature of this equality is that the rhs is not explicitly symmetric under interchange of the three Bessel functions, for example swapping $\mu$ and $\nu$ along with $b$ and $c$.

To cast (9.6) into the form of (9.7), we make use of the second equality in (4.5), and use

$$\sin(x) = \sqrt{\frac{\pi x}{2}} J_{1/2}(x)$$
$$\cos(x) = \sqrt{\frac{\pi x}{2}} J_{-1/2}(x) \qquad (9.8)$$

to write the sin and cos functions as Bessel functions.

If we take different observation times into account, $\boldsymbol{J}$ and $\boldsymbol{M}$ are no longer symmetric (in either the Legendre or Fourier basis). Therefore the definition of $\boldsymbol{D}$ (timing-residual equivalent of Eq. (20) in [9]) changes: the entries of $\boldsymbol{D}$ no longer get symmetrized. As a consequence, $\boldsymbol{D}$ becomes invertible if the pulsar positions are generic, and the pseudoinverse $\boldsymbol{D}^{-1}$ becomes the standard inverse. Note that, as in [9], we use $a < b$ and $c < d$ for pulsar pairs $ab, cd \in \gamma$. Note also that $J^{\mu\nu}$ is no longer required to vanish for $\mu + \nu$ odd.

## X. TRANSMISSION FUNCTIONS

In this section, we calculate the covariance between timing residuals after the universal terms have been removed by fitting a timing model. As a special case, we recover "transmission functions" derived numerically in [10] and analytically in [11].

Before fitting a timing model, the covariance between timing residuals $\mathcal{T}_a(t)$ and $\mathcal{T}_b(t')$ with $t \in [\Delta_a - T_a/2, \Delta_a + T_a/2]$ and $t' \in [\Delta_b - T_b/2, \Delta_b + T_b/2]$ is given by

$$\langle \mathcal{T}_a(t)\mathcal{T}_b(t')\rangle = \int \mathrm{d}f \int \mathrm{d}f' \langle \tilde{\mathcal{T}}_a(f)\tilde{\mathcal{T}}_b^*(f')\rangle \mathrm{e}^{2\pi\mathrm{i}(ft - f't')}$$
$$= \int \mathrm{d}f \left[\frac{H(f)}{\pi f^2}\mu_{ab} + \frac{N_a(f)}{4\pi^2 f^2}\delta_{ab}\right] \mathrm{e}^{2\pi\mathrm{i}f(t-t')}, \qquad (10.1)$$

where the first equality follows from (4.3) and the second from (5.3). The rhs is a function only of the time difference $t - t'$, as must be the case for a stationary random process.

Suppose that we now fit a timing model and subtract the universal terms from the timing residuals. As discussed in Sec. IV, with the expansion (9.2), these postsubtraction timing residuals are

$$\check{\mathcal{T}}_a(t) = \sum_{\mu=3}^{\infty} \mathcal{T}_a^{\mu} P_{\mu}(2(t-\Delta_a)/T_a). \qquad (10.2)$$

The covariance between $\check{\mathcal{T}}_a(t)$ and $\check{\mathcal{T}}_b(t')$ is given by

$$\langle \check{\mathcal{T}}_a(t)\check{\mathcal{T}}_b(t')\rangle = \sum_{\mu=3}^{\infty}\sum_{\nu=3}^{\infty} \langle \mathcal{T}_a^{\mu}\mathcal{T}_b^{\nu}\rangle P_{\mu}(2(t-\Delta_a)/T_a) P_{\nu}(2(t'-\Delta_b)/T_b)$$
$$= \int \mathrm{d}f \left[\frac{H(f)}{\pi f^2}\mu_{ab} + \frac{N_a(f)}{4\pi^2 f^2}\delta_{ab}\right] \mathrm{e}^{2\pi\mathrm{i}f(\Delta_a - \Delta_b)} \left(\sum_{\mu=3}^{\infty}(2\mu+1)\mathrm{i}^{\mu} j_{\mu}(\pi f T_a) P_{\mu}(2(t-\Delta_a)/T_a)\right)$$
$$\times \left(\sum_{\nu=3}^{\infty}(2\nu+1)\mathrm{i}^{\nu} j_{\nu}(\pi f T_b) P_{\nu}(2(t'-\Delta_b)/T_b)\right)^* \qquad (10.3)$$

where the first equality follows from (10.2), and the second equality follows from (5.4) and (9.5) after regrouping terms. "We now use the identity (see Eq. (8.511.4) in [17] )

$$\mathrm{e}^{\mathrm{i}\alpha x} = \sum_{\mu=0}^{\infty}(2\mu+1)\mathrm{i}^{\mu} j_{\mu}(\alpha) P_{\mu}(x), \qquad (10.4)$$

which holds for $-1 \le x \le 1$ and $\alpha \in \mathbb{C}$. From (10.4), it follows that

$$\sum_{\mu=3}^{\infty}(2\mu+1)\mathrm{i}^{\mu} j_{\mu}(\pi f T_a) P_{\mu}(2(t-\Delta_a)/T_a)$$
$$= \mathrm{e}^{2\pi\mathrm{i}f(t-\Delta_a)} S_a(f,t), \qquad (10.5)$$

where we defined

$$S_a(f,t) \equiv 1 - \mathrm{e}^{-2\pi\mathrm{i}f(t-\Delta_a)} \sum_{\mu=0}^{2}(2\mu+1)\mathrm{i}^{\mu} j_{\mu}(\pi f T_a)$$
$$\times P_{\mu}(2(t-\Delta_a)/T_a). \qquad (10.6)$$

To get from (10.4) to (10.5) first replace $\alpha$ with $\pi f T_a$ and $x$ with $2(t-\Delta_a)/T_a$. Then, split the sum in the terms $\nu = 0$, 1, 2 and $\nu = 3, 4, \ldots$, isolate the sum over $\nu = 3, 4, \ldots$ and factor out the phase $\mathrm{e}^{2\pi\mathrm{i}f(t-\Delta_a)}$.

Using (10.5) in (10.3) twice, the covariance of the postfit timing residuals is

$$\langle \check{\mathcal{T}}_a(t)\check{\mathcal{T}}_b(t')\rangle = \int \mathrm{d}f\left[\frac{H(f)}{\pi f^2}\mu_{ab} + \frac{N_a(f)}{4\pi^2 f^2}\delta_{ab}\right] \times S_a(f,t)S_b^*(f,t')\mathrm{e}^{2\pi i f(t-t')}. \tag{10.7}$$

Note that $S_a(f,t)S_b^*(f,t')$ is not a function of $t-t'$, hence the covariance $\langle \check{\mathcal{T}}_a(t)\check{\mathcal{T}}_b(t')\rangle$ is not a function of $t-t'$. This shows that the least-squares subtraction of the constant/linear/quadratic timing model terms makes the timing residuals nonstationary. This point was first made in [12], which remarks that "such a fit is not a stationary process," and independently in [13], which remarks that "stationarity breaks down in the fitting process." Both also demonstrate that the subtraction of the universal timing model terms may be implemented via a projection operation.

We define $R_a(f)$ to be the time average of $|S_a(f,t)|^2$ over $t\in[\Delta_a - T_a/2, \Delta_a + T_a/2]$. Using (3.2) and the identity (see Eq. (10.60.12) in [18])

$$\sum_{n=0}^{\infty}(2n+1)j_n^2(z) = 1, \tag{10.8}$$

we can show from (10.5) that

$$R_a(f) \equiv \frac{1}{T_a}\int_{\Delta_a - T_a/2}^{\Delta_a + T_a/2}\mathrm{d}t|S_a(f,t)|^2 = 1 - \sum_{\mu=0}^{2}(2\mu+1)j_\mu^2(\pi f T_a). \tag{10.9}$$

If $|f| \ll 1/T_a$, then $R_a(f) \propto f^6$, as shown in Fig. 2.

As described in [10], one can interpret $R_a(f)$ as a transmission function, which shows how timing model subtraction effectively removes powers at low frequencies. To see why, first consider the mean-squared timing residuals before timing model subtraction, obtained by setting $a=b$ and $t=t'$ in (10.1):

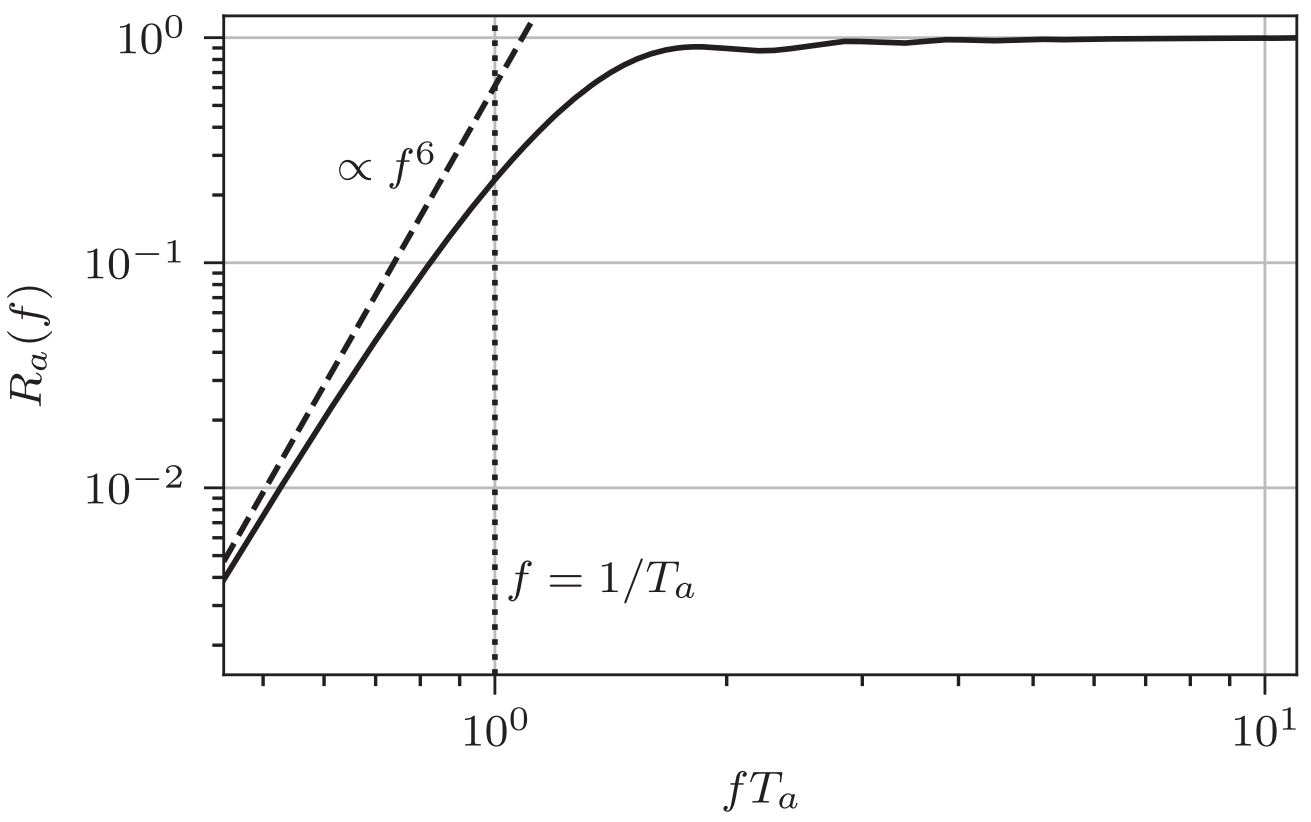

FIG. 2. The transmission function $R_a(f)$ shows low-frequency behavior identical to that of Fig. 2(a) in [10] and Fig. 3 in [11].

$$\langle (\mathcal{T}_a(t))^2\rangle = \int \mathrm{d}f\left[\frac{2H(f)}{3\pi f^2} + \frac{N_a(f)}{4\pi^2 f^2}\right]. \tag{10.10}$$

Note that (as implied by stationarity) this is time independent. In contrast, after the timing model subtraction, and also setting $\Delta_a = 0$, the time average of (10.7) is

$$\frac{1}{T_a}\int_{-T_a/2}^{T_a/2}\mathrm{d}t\langle[\check{\mathcal{T}}_a(t)]^2\rangle = \int \mathrm{d}f\left[\frac{2H(f)}{3\pi f^2} + \frac{N_a(f)}{4\pi^2 f^2}\right]R_a(f). \tag{10.11}$$

The transmission function $R_a(f)$, shown in Fig. 2, is unity at high frequencies but falls off rapidly at frequencies below $1/T_a$. This demonstrates that subtracting the universal timing model terms effectively acts as a filter, removing low-frequency components of both the gravitational-wave background and the pulsar noise. The transmission function is the response of that filter. Equation (10.11), with (10.9) inserted, was also obtained in Eqs. (119) and (120) in [11].

Note that (10.11) can be evaluated analytically for power law $H(f)$ (6.3) and $N_a(f)$ (6.4) if $\gamma < 7$, provided that we impose a high-frequency cutoff, restricting the integral to $|f| \le f_\mathrm{m}$. This gives

$$\frac{1}{T_a}\int_{-T_a/2}^{T_a/2}\mathrm{d}t\langle[\check{\mathcal{T}}_a(t)]^2\rangle = D_2(A_\mathrm{gw}^2 f_\mathrm{r}^{-3}, \gamma_\mathrm{gw}, f_\mathrm{m}, f_\mathrm{r}, T_a) + D_2(A_\mathrm{rn}^2 f_\mathrm{r}^{-3}, \gamma_\mathrm{rn}, f_\mathrm{m}, f_\mathrm{r}, T_a), \tag{10.12}$$

with $D_2$ defined in Appendix B. The limit $f_\mathrm{m} \to \infty$ exists for $1 < \gamma < 7$.

## XI. CONCLUSION

In this paper, we present a systematic development of Legendre polynomials as a basis for PTA data analysis, extending their original application in [11–13]. This basis encodes the universal components of pulsar timing models—namely the constant, linear, and quadratic terms—in the coefficients of the first three Legendre polynomials. The effects of pulsar timing model fitting can be accounted for by setting these three coefficients to zero or, equivalently, by removing those three Legendre polynomials from the basis. We show how this formalism can be used to compute the covariance between coefficients, which can be obtained in closed form, even for bases adapted to pulsars with different observation spans. Finally, we demonstrate how to transform between the Fourier and Legendre bases.

If one uses a complete basis, any scalar quantity should be independent of the basis. For example, Allen and Romano [9] calculated the variance $\sigma_{\hat{\mu}}^2$ of the HD estimator, which may be expressed in terms of a scalar of the form $\boldsymbol{v}^t\boldsymbol{C}^{-1}\boldsymbol{v}$, where $\boldsymbol{v}$ is a column vector and $\boldsymbol{C}$ is a covariance matrix. This scalar quantity is basis independent, so if we compute $\sigma^2_{\hat{\mu}}$ using the Legendre basis (3.1), we should recover the same value as in the Fourier basis (2.1). On the other hand, if we drop terms with $\nu < 3$ in the Legendre basis, this corresponds precisely to a projection onto the modes which remain after pulsar timing subtraction. Dropping these three terms from the summations reduces the number of degrees of freedom by three.

This reduction of dimension is accomplished in the Bayesian analysis by marginalizing over pulsar timing models. Here, this reduction of dimension/projection onto a subspace/elimination of degrees of freedom is accomplished by beginning the summations over basis functions at $\mu, \nu, \lambda, \kappa = 3$. This corresponds exactly to the Allen and Romano analysis, which explicitly states (text after Eq. (16) of [9]) that if the determinants of $\boldsymbol{H}$ or $\boldsymbol{P}_a$ vanish, then their inverses are to be replaced by the Moore-Penrose pseudoinverses.

The reduction of dimension can also be carried out in the Fourier basis. However, in that basis, the linear-in-$t$ and quadratic-in-$t$ terms are represented by an infinite sum of basis elements. In contrast, in the Legendre basis, those terms are completely included in the $\nu = 0$, 1, and 2 modes. So, if a statistic is computed using a finite number of modes, then the Fourier basis does not remove these terms completely.

In this work, our main application is to estimate the HD correlation $\hat{\mu}$ in "position space," meaning, as a function of the interpulsar angle $\gamma$. One can also represent the HD correlation in "harmonic space," as $\mu(\gamma) = \sum_l c_l P_l(\cos\gamma)$, and estimate the coefficients $c_l$. Such estimation was studied in [19], and optimal estimators $\hat{c}_l$ are constructed in [11,20]. These optimal estimators are scalars under basis transformations, so the methods of this paper are applicable and provide a simple way to account for the effects of removing the universal timing terms from pulsar timing residuals.

Our main motivation for this work was to construct tools to evaluate the optimal estimator of the HD correlation and its variance as given in [9]. That evaluation is currently in progress, using public data from the different PTA collaborations [21]. We were led to the use of a Legendre polynomial basis because it provides the simplest way to account for the most important effects of the fitting process that takes place when pulsars are timed and their timing residuals are determined.

It is not yet clear if the Legendre polynomial basis is more broadly useful, though its previous appearance in the literature suggests that this might be the case. As described above, a finite Legendre basis is not equivalent to a finite Fourier basis. For example, the latter cannot exactly represent the linear-in-time and quadratic-in-time terms that are removed by pulsar timing. Thus, comparing equivalent quantities in the two bases gives a way to understand and quantify the effects of the truncation that takes place in the Fourier domain [21]. The use of a Legendre polynomial basis also suggests simple noise spectral models as alternatives to standard ones. For example, a random process might be modeled with coefficients whose amplitudes fall off with increasing Legendre index $\mu$ as a power law in $\mu$.

## ACKNOWLEDGMENTS

We thank Rutger van Haasteren for a number of useful conversations about this topic and, particularly, for emphasizing the effects that fitting pulsar timing models can have on the covariance, Joseph Romano for helpful comments and corrections, and K. J. Lee for calling [12] to our attention. K. D. O was supported in part by NSF Grant No. 2207267 and is a member of the NANOGrav Collaboration, which receives support from NSF Physics Frontiers Center Award No. 2020265.

## DATA AVAILABILITY

No data were created or analyzed in this study.

## APPENDIX A: FITTING OF THE "UNIVERSAL TERMS" IN THE FOURIER BASIS

In the main body of this work we used the Legendre basis to account for the fitting of the "universal terms," the terms constant, linear, and quadratic in $t$, of the timing residuals. In this appendix, we show how to account for the timing model subtractions in the Fourier basis.

In Sec. VII, we determined the basis transformation $\boldsymbol{\Lambda}$. Using $\boldsymbol{\Lambda}$, we can calculate the Fourier basis components of the projection operator $Q$. From (4.1) and (7.2), we have

$$\check{\mathcal{T}}^{\mu}_{a} = Q^{\mu}_{\nu}\mathcal{T}^{\nu}_{a} = Q^{\mu}_{\nu}\Lambda^{\nu}{}_{k}\mathcal{T}^{k}_{a} \tag{A1}$$

with the Einstein summation convention. Relating the projected components in the two bases implies that

$$\check{\mathcal{T}}^{\mu}_{a} = \Lambda^{\mu}{}_{j}\check{\mathcal{T}}^{j}_{a} = \Lambda^{\mu}{}_{j}Q^{j}_{k}\mathcal{T}^{k}_{a}. \tag{A2}$$

Setting the rhs of (A2) equal to the rhs of (A1) and contracting with $(\Lambda^{-1})^{\ell}{}_{\mu}$ gives

$$Q^{\ell}_{k} = (\Lambda^{-1})^{\ell}{}_{\mu}Q^{\mu}_{\nu}\Lambda^{\nu}{}_{k}. \tag{A3}$$

In the Legendre basis, the components of $Q$ (4.2) may be written as

$$Q^{\mu}_{\nu} = \delta^{\mu}_{\nu} - \delta^{\mu}_{0}\delta^{0}_{\nu} - \delta^{\mu}_{1}\delta^{1}_{\nu} - \delta^{\mu}_{2}\delta^{2}_{\nu}. \tag{A4}$$

To evaluate (A3), we need the components of $\boldsymbol{\Lambda}$. From (7.1) and (7.6) these are

$$
\begin{aligned}
{\Lambda^0}_k &= j_0(\pi k)\\
{\Lambda^1}_k &= 3\mathrm{i}j_1(\pi k)\\
{\Lambda^2}_k &= -5j_2(\pi k)\\
{(\Lambda^{-1})^\ell}_0 &= j_0(\pi \ell)\\
{(\Lambda^{-1})^\ell}_1 &= -\mathrm{i}j_1(\pi \ell)\\
{(\Lambda^{-1})^\ell}_2 &= -j_2(\pi \ell).
\end{aligned} \tag{A5}
$$

Putting together (A3)–(A5), the components of $Q$ in the Fourier basis are

$$
Q_k^\ell = \delta_k^\ell - j_0(\pi\ell)j_0(\pi k) - 3j_1(\pi\ell)j_1(\pi k) - 5j_2(\pi\ell)j_2(\pi k). \tag{A6}
$$

Note that these are nonzero for all values of $k$ and $\ell$. This contrasts with the components of $Q$ in the Legendre basis (A4), which vanish off the diagonal $\mu = \nu$ and are either zero or one along the diagonal.

To see how the projection $Q$ modifies $J^{jk}$ and $M_a^{jk}$, consider the Fourier components of the timing residuals $\mathcal{T}_a(t)$, which can be expressed as (see Eq. (13) in [9] for timing residuals)

$$
\mathcal{T}_a^j = \int \mathrm{d}f\, \tilde{\mathcal{T}}_a(f)\mathrm{sinc}\big(\pi(f-f_j)T\big). \tag{A7}
$$

Using (A1), (A6), and (A7), the Fourier components of the projected timing residuals are

$$
\begin{aligned}
\check{\mathcal{T}}_a^j &= Q_k^j \mathcal{T}_a^k\\
&= \mathcal{T}_a^j - \sum_{\ell=0}^{2}(2\ell+1)j_\ell(\pi j)\sum_k j_\ell(\pi k)\mathcal{T}_a^k\\
&= \int \mathrm{d}f\, \tilde{\mathcal{T}}_a(f)\Big[\mathrm{sinc}\big(\pi(f-f_j)T\big)\\
&\quad - \sum_{\ell=0}^{2}(2\ell+1)j_\ell(\pi j)\sum_k j_\ell(\pi k)\mathrm{sinc}\big(\pi(f-f_k)T\big)\Big],
\end{aligned} \tag{A8}
$$

using (A6) for the second equality and (A7) for the third one. We now use the identity

$$
\sum_k j_\ell(\pi k)\mathrm{sinc}\big(\pi(f-f_k)T\big) = j_\ell(\pi f T), \tag{A9}
$$

which follows from (4.5) and the completeness of the Fourier basis. Inserting (A9) into (A8), the Fourier coefficients of the projected timing residuals $\check{\mathcal{T}}_a$ are [cf. Eqs. (126) and (127) in [11] ]

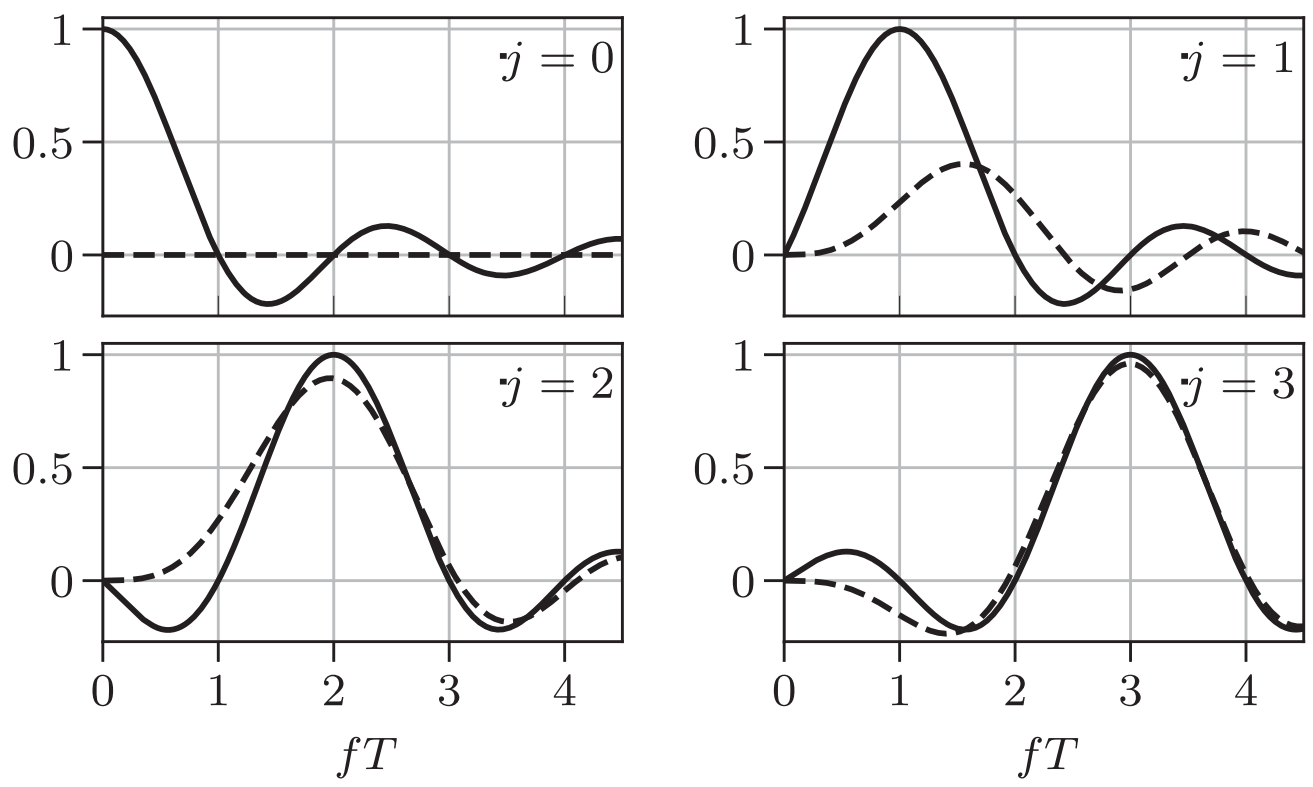

FIG. 3. Comparison of $\mathrm{sinc}(\pi(f-f_j)T)$ (solid lines) with $V_j(f)$ (A11) (dashed lines) for $j = 0,\ldots,3$. The discrepancy decreases as $j$ and $fT$ increase.

$$
\check{\mathcal{T}}_a^j = \int \mathrm{d}f\, \tilde{\mathcal{T}}_a(f)\Big[\mathrm{sinc}\big(\pi(f-f_j)T\big) - \sum_{\ell=0}^{2}(2\ell+1)j_\ell(\pi j)j_\ell(\pi f T)\Big]. \tag{A10}
$$

Motivated by this, we define transfer functions

$$
V_j(f) \equiv \mathrm{sinc}\big(\pi(f-f_j)T\big) - \sum_{\ell=0}^{2}(2\ell+1)j_\ell(\pi j)j_\ell(\pi f T), \tag{A11}
$$

which are illustrated in Fig. 3.

Using these transfer functions, the effects of subtracting the universal timing terms may then be easily expressed in the Fourier basis. These effects may be accounted for by replacing $J^{jk}$ and $M_a^{jk}$ as given in (8.1) with

$$
\begin{aligned}
J^{jk} &\mapsto 4\pi \int \mathrm{d}f \frac{H(f)}{4\pi^2 f^2} V_j(f)V_k(f)\\
M_a^{jk} &\mapsto \int \mathrm{d}f \frac{N_a(f)}{4\pi^2 f^2} V_j(f)V_k(f),
\end{aligned} \tag{A12}
$$

then proceeding as in [9], for example, arriving at expressions for the variance $\sigma_{\hat{\mu}}^2$ in (8.2) and for the squared SNR $\rho^2$ in (8.3).

Note that for the power law $H(f)$ and $N_a(f)$ employed in Sec. VI, the integrals in (A12) have the same range of convergence in $\gamma$ as $J^{\mu\nu}$ and $M_a^{\mu\nu}$ in (5.6). As with (5.6) for $\mu, \nu \geq 3$, they converge if $-1 < \gamma < 7$.

## APPENDIX B: TRANSMISSION FUNCTION INTEGRAL

We calculate the integral

$$D_k(A^2,\gamma,f_{\rm m},f_{\rm r},T)\equiv\int_0^{f_{\rm m}}{\rm d}f\frac{A^2}{12\pi^2}\left(\frac{f}{f_{\rm r}}\right)^{-\gamma}\times\left(1-\sum_{\nu=0}^{k}(2\nu+1)j_\nu^2(\pi fT)\right) \tag{B1}$$

for $k=2$ and $\gamma<7$ using *Mathematica* in two steps. First, we determine the antiderivative of the integrand in (B1) using the `Integrate` method. The antiderivative has a removable singularity at $f=0$. We can determine the leading order term as $f\to0$ using the `Series` method and subtract it from the antiderivative evaluated at $0\le f_{\rm m}<\infty$ to determine the definite integral (B1). For $k=2$ and $\gamma<7$ we obtain

$$\begin{aligned}D_2(A^2,\gamma,f_{\rm m},f_{\rm r},T)=\frac{A^2}{12\pi^2}\Bigg[&-3(2\pi f_{\rm r}T)^{\gamma-1}\pi f_{\rm r}\frac{(\gamma-5)(\gamma-3)(\gamma+2)(\gamma+4)}{\cos(\pi\gamma/2)\Gamma(\gamma+6)}+\left(\frac{f_{\rm m}}{f_{\rm r}}\right)^{-\gamma}\\&\times\Bigg(-\frac{f_{\rm m}}{\gamma-1}+\frac{9f_{\rm m}}{2(\pi f_{\rm m}T)^2(\gamma+1)}+\frac{9f_{\rm m}}{(\pi f_{\rm m}T)^4(\gamma+3)}+\frac{45f_{\rm m}}{2(\pi f_{\rm m}T)^6(\gamma+5)}\\&+\frac{f_{\rm m}}{64(\pi f_{\rm m}T)^6}\Big[6\Big((\gamma-6)(4+\gamma(\gamma-1))-4(\pi f_{\rm m}T)^2(\gamma-14)\Big)\cos(2\pi f_{\rm m}T)\\&-6(\gamma-5)(\gamma-3)(\gamma+2)(\gamma+4)(2\pi f_{\rm m}T)^{\gamma+5}{\rm ci}(-\gamma-5,2\pi f_{\rm m}T)\\&+12\pi f_{\rm m}T(4(\pi f_{\rm m}T)^2-54-\gamma(\gamma-11))\sin(2\pi f_{\rm m}T)\Big]\Bigg)\Bigg],\end{aligned} \tag{B2}$$

where the generalized cosine integral is ${\rm ci}(a,z)\equiv\int_z^\infty t^{a-1}\cos t\,{\rm d}t$. The limit $f_{\rm m}\to\infty$ exists for $1<\gamma<7$ and is given by

$$\lim_{f_{\rm m}\to\infty}D_2(A^2,\gamma,f_{\rm m},f_{\rm r},T)=-3\frac{A^2}{12\pi^2}(2\pi f_{\rm r}T)^{\gamma-1}\pi f_{\rm r}\frac{(\gamma-5)(\gamma-3)(\gamma+2)(\gamma+4)}{\cos(\pi\gamma/2)\Gamma(\gamma+6)}. \tag{B3}$$

An identical result is obtained in [12]. Starting from Eq. (A11) in [12], setting $S_0\to A^2f_{\rm r}^\gamma/12\pi^2$, $\beta\to\gamma$, and $n\to2$, then using the recurrence and reflection formulas for the Γ-function, one obtains $(\sigma')^2=D_2(A^2,\gamma,\infty,f_{\rm r},T)$ as given in (B3).

## APPENDIX C: REDSHIFTS

Some of the literature, for example [9], describes arrival time perturbations in terms of pulsar "redshifts." These redshifts are the derivatives of the timing residuals

$$Z_a(t)=\frac{\rm d}{{\rm d}t}\mathcal{T}_a(t). \tag{C1}$$

A positive redshift means that the pulsar rotation frequency (typically hundred of Hz) decreases slightly: the pulses come farther and farther apart, $\mathcal{T}_a(t)$ increases with time $t$, and $Z_a(t)>0$.

Below, we discuss how an analysis carried out in terms of timing residuals could alternatively be done using redshifts, and how redshifts are affected by the removal of the universal timing terms.

The term "redshift" is used figuratively rather than literally, because the frequency shifts in question are not of visible light. Rather, they refer to minuscule changes in the pulsar rotation frequency, as observed at the SSB. The pulsar's rotation makes it act as a clock, and by the principle of equivalence, all clocks are affected by gravity in the same way. So even though a typical pulsar's pulsation frequency is 12 orders of magnitude smaller than for visible light, the fractional frequency change produced by a gravitational wave is identical. Thus, the literature uses the word "redshift" for both.

If $\lambda$ denotes the spatial distance between successive pulses arriving at the SSB, and the differences from the mean are $\delta\lambda$, then the redshift is $Z=\delta\lambda/\lambda=-\delta f/f$ where $f$ is the mean pulsar rotation frequency [22] and $\delta f$ is the difference from the mean.

In the same way as for timing residuals, we can expand the time-dependent redshift (C1) of pulsar $a$ in a Legendre polynomial basis, as

$$Z_a(t)=\sum_\mu Z_a^\mu P_\mu(2t/T)\quad\text{for } t\in[-T/2,T/2]. \tag{C2}$$

The relationship of the Legendre coefficients $Z_a^\mu$ to the Fourier transform of the redshift is given by the analog of (4.6):

$$Z_a^\mu=(2\mu+1){\rm i}^\mu\int{\rm d}f\,\tilde{Z}_a(f)j_\mu(\pi fT). \tag{C3}$$

Recall from (5.1) that the covariance of these redshift Fourier amplitudes is

$$\langle \tilde{Z}_a(f)\tilde{Z}_b^*(f')\rangle = \left[4\pi H(f)\mu_{ab} + N_a(f)\delta_{ab}\right]\delta(f-f'). \tag{C4}$$

So, in the same way that (5.4) was derived, we find that the covariance of the redshift Legendre components is

$$\begin{aligned}\langle Z_a^\mu Z_b^\nu\rangle &= (2\mu+1)(2\nu+1)\mathrm{i}^{\mu-\nu}\int \mathrm{d}f \int \mathrm{d}f' \langle \tilde{Z}_a(f)\tilde{Z}^*_b(f')\rangle j_\mu(\pi f T) j_\nu(\pi f' T)\\ &= H^{\mu\nu}\mu_{ab} + N_a^{\mu\nu}\delta_{ab},\end{aligned} \tag{C5}$$

where the covariance matrices are

$$\begin{aligned}H^{\mu\nu} &\equiv 4\pi(2\mu+1)(2\nu+1)\mathrm{i}^{\mu-\nu}\int \mathrm{d}f\, H(f) j_\mu(\pi f T) j_\nu(\pi f T)\\ N_a^{\mu\nu} &\equiv (2\mu+1)(2\nu+1)\mathrm{i}^{\mu-\nu}\int \mathrm{d}f\, N_a(f) j_\mu(\pi f T) j_\nu(\pi f T).\end{aligned} \tag{C6}$$

These should be compared to (5.4) and (5.5); they differ by an overall factor of $4\pi^2 f^2$ inside the integrals.

We now proceed to recast a timing-residual analysis in terms of the redshifts $Z_a(t)$. We start by expressing the Legendre components $Z_a^\mu$ of the redshifts in terms of the Legendre components $\mathcal{T}_a^\mu$ of the timing residuals. The relation (C1) implies (5.2), so

$$\tilde{Z}_a(f) = 2\pi\mathrm{i}f\tilde{\mathcal{T}}_a(f). \tag{C7}$$

Inserting (C7) into (C3) gives

$$Z_a^\mu = \frac{2}{T}(2\mu+1)\mathrm{i}^{\mu+1}\int \mathrm{d}f\, \pi f T j_\mu(\pi f T)\tilde{\mathcal{T}}(f). \tag{C8}$$

Spherical Bessel functions satisfy the recurrence relation

$$z j_\mu(z) = [2(\mu+1)+1] j_{\mu+1}(z) - z j_{\mu+2}(z). \tag{C9}$$

Applying the relation (C9) iteratively, we obtain

$$\begin{aligned}z j_\mu(z) = &[2(\mu+1)+1] j_{\mu+1}(z)\\ &- [2(\mu+3)+1] j_{\mu+3}(z)\\ &+ [2(\mu+5)+1] j_{\mu+5}(z)\\ &- \ldots.\end{aligned} \tag{C10}$$

Take $z = \pi f T$ in (C10), insert it into (C8), and use (4.6). This gives

$$\begin{aligned}Z_a^\mu &= \frac{2}{T}(2\mu+1)\left[\mathcal{T}_a^{\mu+1} + \mathcal{T}_a^{\mu+3} + \mathcal{T}_a^{\mu+5} + \ldots\right]\\ &= A^\mu{}_\nu \mathcal{T}_a^\nu,\end{aligned} \tag{C11}$$

where the summation convention is used after the second equality and in what follows. Here,

$$\boldsymbol{A} \equiv A^\mu{}_\nu \equiv \frac{2}{T}\begin{pmatrix} 0 & 1 & 0 & 1 & 0 & 1 & 0 & \ldots\\ 0 & 0 & 3 & 0 & 3 & 0 & 3 & \ldots\\ 0 & 0 & 0 & 5 & 0 & 5 & 0 & \ldots\\ 0 & 0 & 0 & 0 & 7 & 0 & 7 & \ldots\\ 0 & 0 & 0 & 0 & 0 & 9 & 0 & \ldots\\ 0 & 0 & 0 & 0 & 0 & 0 & 11 & \ldots\\ \vdots & \vdots & \vdots & \vdots & \vdots & \vdots & \vdots & \ddots \end{pmatrix} \tag{C12}$$

is a matrix with rows indexed by $\mu$ and columns indexed by $\nu$. Thus, to obtain redshifts from timing residuals, apply the (noninvertible) linear operator $\boldsymbol{A}$.

We also define a linear projection operator $P$ which removes terms in the redshift which are constant and linear in $t$. This is similar to $Q$ defined in Sec. IV for timing residuals. As before, such projected quantities are indicated with a háček accent ˇ. The projected redshifts are

$$\check{Z}_a(t) \equiv (PZ_a)(t), \tag{C13}$$

where the components of $P$ in the Legendre basis are

$$P^\mu_\nu = \begin{cases} \delta^\mu_\nu & \text{for } 2 \le \nu \text{ and } 2 \le \mu\\ 0 & \text{otherwise.}\end{cases} \tag{C14}$$

With this, we can define another useful operator, similar to $\boldsymbol{A}$.

The linear operator $\check{\boldsymbol{A}}$ which produces projected redshifts from timing residuals follows from substituting (C11) into (C13). We obtain

$$\begin{aligned}\check{Z}^\mu_a &= \check{A}^\mu{}_\nu \mathcal{T}_a^\nu\\ &= \begin{cases} 0 & \text{for } \mu \le 1\\ \frac{2}{T}(2\mu+1)\sum_\nu \mathcal{T}_a^{\mu+1+2\nu} & \text{otherwise,}\end{cases}\end{aligned} \tag{C15}$$

where

$$\check{\boldsymbol{A}} \equiv \check{A}^{\mu}{}_{\nu} \equiv P^{\mu}_{\lambda} A^{\lambda}{}_{\nu} = \frac{2}{T} \begin{pmatrix} 0 & 0 & 0 & 0 & 0 & 0 & 0 & \dots \\ 0 & 0 & 0 & 0 & 0 & 0 & 0 & \dots \\ 0 & 0 & 0 & 5 & 0 & 5 & 0 & \dots \\ 0 & 0 & 0 & 0 & 7 & 0 & 7 & \dots \\ 0 & 0 & 0 & 0 & 0 & 9 & 0 & \dots \\ 0 & 0 & 0 & 0 & 0 & 0 & 11 & \dots \\ \vdots & \vdots & \vdots & \vdots & \vdots & \vdots & \vdots & \ddots \end{pmatrix}. \quad \text{(C16)}$$

This should be compared with (C12): The operator $\boldsymbol{A}$ has a one-dimensional null space, whereas $\check{\boldsymbol{A}}$ has a three-dimensional null space.

Identical projected redshifts are obtained starting from either the original timing residuals or from the projected timing residuals, since

$$\check{Z}^{\mu}_{a} = \check{A}^{\mu}_{\nu} \mathcal{T}^{\nu}_{a} = \check{A}^{\mu}_{\nu} Q^{\nu}_{\lambda} \mathcal{T}^{\lambda}_{a} = \check{A}^{\mu}_{\nu} \check{\mathcal{T}}^{\nu}_{a}. \quad \text{(C17)}$$

This follows because the projection $Q$ sets the components $\mathcal{T}^0_a$, $\mathcal{T}^1_a$, and $\mathcal{T}^2_a$ to zero. From (C11), the only components of $Z^{\mu}_a$ where these enter are $Z^0_a$ and $Z^1_a$, and those are precisely the components that the projection $P$ sets to zero in (C17).

An important observation is that the second equality in (C17) would not hold without the projection $P$ hidden inside $\check{\boldsymbol{A}}$, meaning that $\check{Z}^{\mu}_a \neq A^{\mu}{}_{\nu} \check{\mathcal{T}}^{\nu}_a$. This is because, by (C11), $A^{\mu}{}_{\nu} \check{\mathcal{T}}^{\nu}_a$ has nonzero $Z^0_a$ and $Z^1_a$. This corresponds to the following inequality of operations applied to timing residuals:

$$P \circ \frac{\mathrm{d}}{\mathrm{d}t} = P \circ \frac{\mathrm{d}}{\mathrm{d}t} \circ Q \neq \frac{\mathrm{d}}{\mathrm{d}t} \circ Q. \quad \text{(C18)}$$

A simple example is to take $\mathcal{T}_a(t) = P_3(2t/T)$, corresponding to $\mathcal{T}^{\mu}_a = \delta^{\mu}_3$. Then, by (4.2), $\check{\mathcal{T}}^{\mu}_a = \mathcal{T}^{\mu}_a = \delta^{\mu}_3$. The expression on the far left of (C17) and (C18) is

$$\check{Z}^{\mu}_{a} = \left[\left(P \circ \frac{\mathrm{d}}{\mathrm{d}t}\right) \mathcal{T}_a\right]^{\mu} = \check{A}^{\mu}{}_{\nu} \mathcal{T}^{\nu}_a = \frac{10}{T} \delta^{\mu}_2, \quad \text{(C19)}$$

where we used (C16). The expression on the far right of (C18) is

$$\left[\left(\frac{\mathrm{d}}{\mathrm{d}t} \circ Q\right) \mathcal{T}_a\right]^{\mu} = A^{\mu}{}_{\nu} \check{\mathcal{T}}^{\nu}_a = \frac{2}{T} \delta^{\mu}_0 + \frac{10}{T} \delta^{\mu}_2, \quad \text{(C20)}$$

where we used (C12). The fact that (C19) and (C20) are different illustrates the inequality of (C18).

The inequality in (C18) is because the projection operation which minimizes the mean square of the redshift by removing constant and linear terms is not the same as the projection operation which minimizes the mean square of the timing residuals by removing constant, linear, and quadratic terms. This is directly visible in the first equality of (C11): even if $\mathcal{T}^0_a$, $\mathcal{T}^1_a$, and $\mathcal{T}^2_a$ all vanish, one might have $Z^0_a$ and/or $Z^1_a$ nonzero.

In Appendix D, we need the Moore-Penrose pseudoinverse of $\check{\boldsymbol{A}}$. The kernel of $\check{\boldsymbol{A}}$ is the three-dimensional vector space

$$\ker \check{\boldsymbol{A}} = u \begin{pmatrix} 1 \\ 0 \\ 0 \\ 0 \\ \vdots \\ 0 \end{pmatrix} + v \begin{pmatrix} 0 \\ 1 \\ 0 \\ 0 \\ \vdots \\ 0 \end{pmatrix} + w \begin{pmatrix} 0 \\ 0 \\ 1 \\ 0 \\ \vdots \\ 0 \end{pmatrix}, \quad \text{(C21)}$$

for arbitrary real numbers $u$, $v$, $w$. Let $\check{\boldsymbol{A}}^{-1}$ denote the pseudoinverse of $\check{\boldsymbol{A}}$. Then by definition $\check{\boldsymbol{A}}^{-1}\check{\boldsymbol{A}}$ is a projector onto the $(N-3)$-dimensional complement of $\ker \check{\boldsymbol{A}}$. This is also the image of $\boldsymbol{Q}$, hence $\check{\boldsymbol{A}}^{-1}\check{\boldsymbol{A}} = \boldsymbol{Q}$. Furthermore, by definition $\check{\boldsymbol{A}}\check{\boldsymbol{A}}^{-1}$ is a projector onto the image of $\check{\boldsymbol{A}}$. From (C16) we see that the image of $\check{\boldsymbol{A}}$ is also the image of $\boldsymbol{P}$, hence $\check{\boldsymbol{A}}\check{\boldsymbol{A}}^{-1} = \boldsymbol{P}$.

## APPENDIX D: THE OPTIMAL HD ESTIMATOR USING REDSHIFTS

We assume that the sum over Legendre indices in (3.1) terminates at some order $N$. This is needed to ensure the existence of the necessary Moore-Penrose pseudoinverses, and holds in any practical implementation, where timing residuals are reconstructed from finitely many data points. Thus, the linear operators $\boldsymbol{P} \equiv P^{\mu}_{\nu}$, $\boldsymbol{A} \equiv A^{\mu}{}_{\nu}$, and $\boldsymbol{Q} \equiv Q^{\mu}_{\nu}$ become finite-dimensional matrices.

In Sec. VIII, starting from timing residuals, we showed that the variance of the optimal HD estimator is given by (8.9), and that the squared expected SNR in that angular bin is given by (8.10). As discussed in Sec. VIII A, for postfit timing residuals, we replace $\boldsymbol{J}$ and $\boldsymbol{D}$ by $\check{\boldsymbol{J}}$ and $\check{\boldsymbol{D}}$, defined in (8.11). Thus, the variance and squared expected SNR calculated from postfit timing residuals are

$$\check{\sigma}^2_{\hat{\mu}} \equiv \frac{\mu^2_{\mathrm{u}}(\gamma)}{\sum_{ab\in\gamma} \sum_{cd\in\gamma} \sum_{\mu,\nu,\lambda,\kappa} \mu_{ab} \check{J}^{\mu\nu} (\check{D}^{-1})^{ab,cd}_{\mu\nu,\lambda\kappa} \mu_{cd} \check{J}^{\lambda\kappa}}, \quad \text{(D1)}$$

and

$$\check{\rho}^2 \equiv \sum_{ab\in\gamma} \sum_{cd\in\gamma} \sum_{\mu,\nu,\lambda,\kappa} \mu_{ab} \check{J}^{\mu\nu} (\check{D}^{-1}_0)^{ab,cd}_{\mu\nu,\lambda\kappa} \mu_{cd} \check{J}^{\lambda\kappa}. \quad \text{(D2)}$$

In what follows, we show that the same $\check{\sigma}^2_{\hat{\mu}}$ and $\check{\rho}^2$ are obtained starting from redshifts, using, unexpectedly, *any* of the three options in (C18), see Fig. 4.

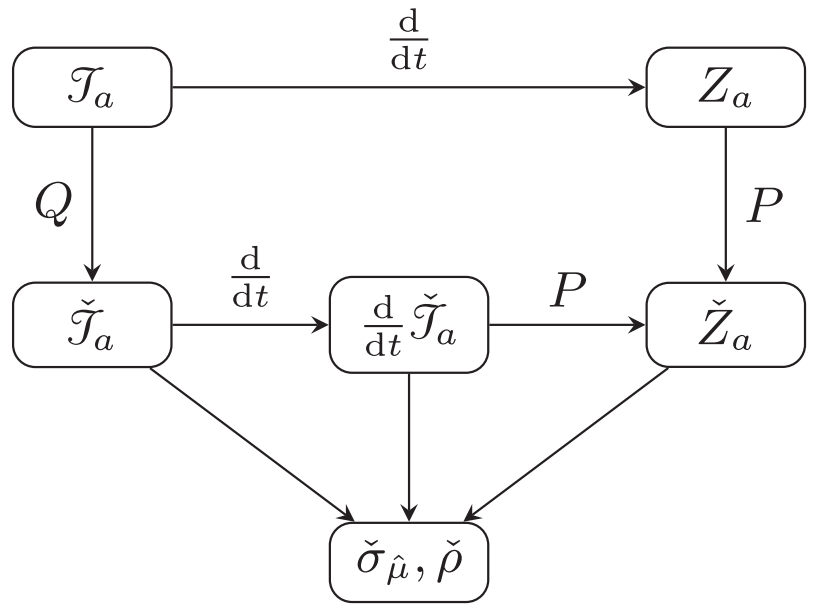

FIG. 4. Starting from timing residuals $\mathcal{T}_a$ in the top left corner, we always end up with the same $\check{\sigma}_{\hat{\mu}}$ and $\check{\rho}$. This is independent of the choice of Fourier or Legendre basis to evaluate $\check{\sigma}_{\hat{\mu}}$ and $\check{\rho}$.

*Option 1.* We start with $P \circ \mathrm{d}/\mathrm{d}t$ applied to timing residuals $\mathcal{T}_a(t)$, giving projected redshifts $\check{Z}_a(t)$ with Legendre coefficients $\check{Z}^{\mu}_a = \check{A}^{\mu}{}_{\nu}\mathcal{T}^{\nu}_a$. Following the same logic as in [9], the variance of the optimal HD estimator and SNR are given by

$$\check{\sigma}^2_{\hat{\mu}} = \frac{\mu^2_{\mathrm{u}}(\gamma)}{\sum_{ab\in\gamma}\sum_{cd\in\gamma}\sum_{\mu,\nu,\lambda,\kappa}\mu_{ab}\check{K}^{\mu\nu}(\check{C}^{-1})^{ab,cd}_{\mu\nu,\lambda\kappa}\mu_{cd}\check{K}^{\lambda\kappa}} \tag{D3}$$

and

$$\check{\rho}^2 = \sum_{ab\in\gamma}\sum_{cd\in\gamma}\sum_{\mu,\nu,\lambda,\kappa}\mu_{ab}\check{K}^{\mu\nu}(\check{C}^{-1}_0)^{ab,cd}_{\mu\nu,\lambda\kappa}\mu_{cd}\check{K}^{\lambda\kappa}. \tag{D4}$$

Here, $\check{K}^{\mu\nu}$ is the covariance matrix of projected redshift correlations arising from the GWB, and $\check{C}^{\mu\nu,\lambda\kappa}_{ab,cd}$ is the covariance matrix of projected redshift products. We have used the same symbols on the lhs of (D3) and (D4) as in (D1) and (D2), because they are equal, as we will now show.

The projected redshift covariance is given by (5.4) and (C15) as

$$\langle \check{Z}^{\mu}_a \check{Z}^{\nu}_b \rangle = \check{K}^{\mu\nu}\mu_{ab} + \check{A}^{\mu}{}_{\lambda}M^{\lambda\kappa}_a(\check{A}^t)_{\kappa}{}^{\nu}\delta_{ab}. \tag{D5}$$

Here, $(\check{A}^t)_{\kappa}{}^{\nu} \equiv \check{A}^{\nu}{}_{\kappa}$ and

$$\check{\boldsymbol{K}} \equiv \check{K}^{\mu\nu} \equiv \check{A}^{\mu}{}_{\lambda}J^{\lambda\kappa}(\check{A}^t)_{\kappa}{}^{\nu} \tag{D6}$$

is the GWB-induced covariance of the projected pulsar redshifts.

The covariance of the product of projected redshifts is defined by analogy with [9] [Eqs. (19) and (20)], as

$$\check{C}^{\mu\nu,\lambda\kappa}_{ab,cd} \equiv \big\langle \check{Z}^{(\mu}_a\check{Z}^{\nu)}_b\check{Z}^{(\lambda}_c\check{Z}^{\kappa)}_d\big\rangle - \big\langle \check{Z}^{(\mu}_a\check{Z}^{\nu)}_b\big\rangle\big\langle\check{Z}^{(\lambda}_c\check{Z}^{\kappa)}_d\big\rangle. \tag{D7}$$

Here, the curved brackets around the indices denote symmetrization as defined in [9]. From (C15), it follows that

$$\check{\boldsymbol{C}} \equiv \check{C}^{\mu\nu,\lambda\kappa}_{ab,cd} = \check{A}^{\mu}{}_{\alpha}\check{A}^{\nu}{}_{\beta}D^{\alpha\beta,\gamma\delta}_{ab,cd}(\check{A}^t)_{\gamma}{}^{\lambda}(\check{A}^t)_{\delta}{}^{\kappa}. \tag{D8}$$

The pseudoinverse of $\check{\boldsymbol{C}}$ is

$$(\check{C}^{-1})^{ab,cd}_{\mu\nu,\lambda\kappa} = (\check{A}^{-t})_{\mu}{}^{\alpha}(\check{A}^{-t})_{\nu}{}^{\beta}(\check{D}^{-1})^{ab,cd}_{\alpha\beta,\gamma\delta}(\check{A}^{-1})^{\gamma}{}_{\lambda}(\check{A}^{-1})^{\delta}{}_{\kappa}, \tag{D9}$$

where $\check{\boldsymbol{A}}^{-t} \equiv (\check{\boldsymbol{A}}^{-1})^t$. To see that this is the pseudoinverse of $\check{\boldsymbol{C}}$, contract (D9) with (D8), and use $\check{\boldsymbol{A}}^{-1}\check{\boldsymbol{A}} = \boldsymbol{Q}$, (8.11) and (8.12). One obtains

$$\begin{aligned}\sum_{cd\in\gamma}\sum_{\lambda,\kappa}(\check{C}^{-1})^{ab,cd}_{\mu\nu,\lambda\kappa}\check{C}^{\lambda\kappa,\alpha\beta}_{cd,ef} &= \delta^a_e\delta^b_f\check{A}^{\alpha}{}_{\lambda}(\check{A}^{-1})^{\lambda}{}_{\mu}\check{A}^{\beta}{}_{\kappa}(\check{A}^{-1})^{\kappa}{}_{\nu} \\ &= \delta^a_e\delta^b_f P^{\alpha}_{\mu}P^{\beta}_{\nu},\end{aligned} \tag{D10}$$

where $a < b$ and $e < f$. In the final equality, we used that $\check{\boldsymbol{A}}\check{\boldsymbol{A}}^{-1} = \boldsymbol{P}$.

Substituting (D6) and (D9) into the denominator of (D3) then shows that $\check{\sigma}^2_{\hat{\mu}}$ as defined by (D3) with projected redshifts is equal to (D1) as defined with postfit timing residuals. The same argument shows that $\check{\rho}^2$ as defined by (D4) is equal to (D2).

*Option 2.* We apply $P \circ \mathrm{d}/\mathrm{d}t \circ Q$ to timing residuals $\mathcal{T}_a(t)$. However, by (C18), $P \circ \mathrm{d}/\mathrm{d}t \circ Q = P \circ \mathrm{d}/\mathrm{d}t$ when applied to timing residuals. Therefore option 2 is equivalent to option 1: both cases result in projected redshifts $\check{Z}_a(t)$, and (D1) equals (D3) and (D2) equals (D4).

*Option 3.* Now we consider $\mathrm{d}/\mathrm{d}t \circ Q$ applied to timing residuals $\mathcal{T}_a(t)$. Recall from (C18) that the resulting function is not $\check{Z}_a(t)$. Therefore there is, *a priori*, no reason why option 3 should be equivalent to options 1 and 2.

However, as shown in Appendix C, the function $(\mathrm{d}/\mathrm{d}t \circ Q)\mathcal{T}_a$ has Legendre coefficients $A^{\mu}{}_{\nu}\check{\mathcal{T}}^{\nu}_a = A^{\mu}{}_{\nu}Q^{\nu}_{\lambda}\mathcal{T}^{\lambda}_a$. We can now proceed analogously to option 1 and calculate the covariance matrix of correlations of Legendre coefficients. This leads to $\check{\boldsymbol{A}}$ being replaced by $\boldsymbol{AQ}$ in (D6), (D8), and (D9). In (D9), this is not obvious, since we do not know the pseudoinverse of $\boldsymbol{AQ}$. From (4.2) and (C12) we obtain

$$\boldsymbol{AQ} = \frac{2}{T}\begin{pmatrix} 0 & 0 & 0 & 1 & 0 & 1 & 0 & \dots \\ 0 & 0 & 0 & 0 & 3 & 0 & 3 & \dots \\ 0 & 0 & 0 & 0 & 0 & 5 & 0 & \dots \\ 0 & 0 & 0 & 0 & 0 & 0 & 7 & \dots \\ \vdots & \vdots & \vdots & \vdots & \vdots & \vdots & \vdots & \ddots \end{pmatrix}. \tag{D11}$$

The kernel of $\boldsymbol{AQ}$ is the same as the kernel of $\check{\boldsymbol{A}}$. Thus, as for $\check{\boldsymbol{A}}$, we obtain $(\boldsymbol{AQ})^{-1}\boldsymbol{AQ} = \boldsymbol{Q}$, where $(\boldsymbol{AQ})^{-1}$ is the Moore-Penrose pseudoinverse of $\boldsymbol{AQ}$. In contrast to $\check{\boldsymbol{A}}$, the image of $\boldsymbol{AQ}$ is the whole space, so $\boldsymbol{AQ}(\boldsymbol{AQ})^{-1} = \mathbb{1}$. Hence, $\check{\boldsymbol{A}}$ may be replaced by $\boldsymbol{AQ}$ in (D9), and in (D10), $P^{\alpha}_{\mu}P^{\beta}_{\nu}$ may be replaced by $\delta^{\alpha}_{\mu}\delta^{\beta}_{\nu}$. This reflects that option 3 does not project with $P$. As with options 1 and 2, we again obtain (D3) and (D4).